%% file: main.tex
\pdfoutput=1
\documentclass[12pt,a4paper]{article}
\usepackage{ifthen} % for conditional statements
\newboolean{pdflatex}
\setboolean{pdflatex}{true} % False for eps figures 

\newboolean{articletitles}
\setboolean{articletitles}{true} % False removes titles in references

\newboolean{uprightparticles}
\setboolean{uprightparticles}{false} %True for upright particle symbols

\newboolean{inbibliography}
\setboolean{inbibliography}{false} %True once you enter the bibliography

\usepackage{longtable} % only for template; not usually to be used in PAPERs
\usepackage{multirow}
\input{preamble}

\begin{document}

%%%%%%%%%%%%%%%%%%%%%%%%%
%%%%% Title     %%%%%%%%%
%%%%%%%%%%%%%%%%%%%%%%%%%
\renewcommand{\thefootnote}{\fnsymbol{footnote}}
\setcounter{footnote}{1}

% %%%%%%% CHOOSE TITLE PAGE--------
%\onecolumn
% \input{title-LHCb-ANA}
%\input{title-LHCb-CONF}
\input{title-LHCb-PAPER}

%\twocolumn
% %%%%%%%%%%%%% ---------

\renewcommand{\thefootnote}{\arabic{footnote}}
\setcounter{footnote}{0}

%%%%%%%%%%%%%%%%%%%%%%%%%%%%%%%%
%%%%%  Table of Content   %%%%%%
%%%%%%%%%%%%%%%%%%%%%%%%%%%%%%%%
%%%% Uncomment next 2 lines if desired
%\tableofcontents
%\cleardoublepage

%%%%%%%%%%%%%%%%%%%%%%%%%
%%%%% Main text %%%%%%%%%
%%%%%%%%%%%%%%%%%%%%%%%%%

\pagestyle{plain} % restore page numbers for the main text
\setcounter{page}{1}
\pagenumbering{arabic}

%% Uncomment during review phase. 
%% Comment before a final submission.
%\linenumbers

% You can include short sections directly in the main tex file.
% However, for larger papers it is desirable to split the text into
% several semiautonomous files, which can be revised independently.
% This is especially useful when developing a document in
% collaboration with several people, since then different parts can be
% edited independently.  This type of file organization is shown here.
% 

\input{body}

% Do not include this in analysis note and conference reports
\input{acknowledgements}

\addcontentsline{toc}{section}{References}
\setboolean{inbibliography}{true}
\bibliographystyle{LHCb}
\bibliography{main,LHCb-PAPER,LHCb-CONF,LHCb-DP}

\end{document}

%% file: preamble.tex
\usepackage{microtype}
\usepackage{lineno}  % for line numbering during review
\usepackage{xspace} % To avoid problems with missing or double spaces after
\usepackage{graphicx}  % to include figures (can also use other packages)
\usepackage{color}
\usepackage{colortbl}
\graphicspath{{./figs/}} % Make Latex search fig subdir for figures

\usepackage{amsmath} % Adds a large collection of math symbols
\usepackage{amssymb}
\usepackage{amsfonts}
\usepackage{upgreek} % Adds in support for greek letters in roman typeset

\newcommand*\patchAmsMathEnvironmentForLineno[1]{%
\expandafter\let\csname old#1\expandafter\endcsname\csname #1\endcsname
\expandafter\let\csname oldend#1\expandafter\endcsname\csname
end#1\endcsname
 \renewenvironment{#1}%
   {\linenomath\csname old#1\endcsname}%
   {\csname oldend#1\endcsname\endlinenomath}%
}
\newcommand*\patchBothAmsMathEnvironmentsForLineno[1]{%
  \patchAmsMathEnvironmentForLineno{#1}%
  \patchAmsMathEnvironmentForLineno{#1*}%
}
\AtBeginDocument{%
\patchBothAmsMathEnvironmentsForLineno{equation}%
\patchBothAmsMathEnvironmentsForLineno{align}%
\patchBothAmsMathEnvironmentsForLineno{flalign}%
\patchBothAmsMathEnvironmentsForLineno{alignat}%
\patchBothAmsMathEnvironmentsForLineno{gather}%
\patchBothAmsMathEnvironmentsForLineno{multline}%
}

\usepackage{hyperref}    % Hyperlinks in references
\usepackage[all]{hypcap} % Internal hyperlinks to floats.

\input{lhcb-symbols-def} % Add in the predefined LHCb symbols

\usepackage{cite} % Allows for ranges in citations
\usepackage{mciteplus}

%% file: lhcb-symbols-def.tex
\def\lhcb {\mbox{LHCb}\xspace}

\def\rich   {RICH\xspace}

\ifthenelse{\boolean{uprightparticles}}%
{

 \def\Ppi         {\ensuremath{\uppi}\xspace}

 \def\PDelta      {\ensuremath{\Delta}\xspace}                 
 \def\PXi      {\ensuremath{\Xi}\xspace}                 
 \def\PLambda      {\ensuremath{\Lambda}\xspace}                 
 \def\PSigma      {\ensuremath{\Sigma}\xspace}                 
 \def\POmega      {\ensuremath{\Omega}\xspace}                 
 \def\PUpsilon      {\ensuremath{\Upsilon}\xspace}                 
 
 \def\PB      {\ensuremath{\mathrm{B}}\xspace}                 
                  
 \def\PD      {\ensuremath{\mathrm{D}}\xspace}

 \def\PK      {\ensuremath{\mathrm{K}}\xspace}

 \def\Pb      {\ensuremath{\mathrm{b}}\xspace}                 
 \def\Pc      {\ensuremath{\mathrm{c}}\xspace}

 \def\Pi      {\ensuremath{\mathrm{i}}\xspace}

 \def\Ps      {\ensuremath{\mathrm{s}}\xspace}

}
{

 \def\Ppi         {\ensuremath{\pi}\xspace}

 \mathchardef\PDelta="7101
 \mathchardef\PXi="7104
 \mathchardef\PLambda="7103
 \mathchardef\PSigma="7106
 \mathchardef\POmega="710A
 \mathchardef\PUpsilon="7107
                  
 \def\PB      {\ensuremath{B}\xspace}                 
                  
 \def\PD      {\ensuremath{D}\xspace}

 \def\PK      {\ensuremath{K}\xspace}

 \def\Pb      {\ensuremath{b}\xspace}                 
 \def\Pc      {\ensuremath{c}\xspace}

 \def\Pi      {\ensuremath{i}\xspace}

 \def\Ps      {\ensuremath{s}\xspace}

}

\def\squark    {\ensuremath{\Ps}\xspace}

\def\cquark    {\ensuremath{\Pc}\xspace}

\def\bquark    {\ensuremath{\Pb}\xspace}

\def\pion  {\ensuremath{\Ppi}\xspace}

\def\pip   {\ensuremath{\pion^+}\xspace}
\def\pim   {\ensuremath{\pion^-}\xspace}

\def\kaon  {\ensuremath{\PK}\xspace}
  \def\Kbar  {\kern 0.2em\overline{\kern -0.2em \PK}{}\xspace}

\def\Kp    {\ensuremath{\kaon^+}\xspace}
\def\Km    {\ensuremath{\kaon^-}\xspace}

  \def\Dbar    {\kern 0.2em\overline{\kern -0.2em \PD}{}\xspace}

\def\B       {\ensuremath{\PB}\xspace}
\def\Bbar    {\ensuremath{\kern 0.18em\overline{\kern -0.18em \PB}{}}\xspace}
\def\Bb      {\ensuremath{\Bbar}\xspace}

\def\Bd      {\ensuremath{\B^0}\xspace}
\def\Bs      {\ensuremath{\B^0_\squark}\xspace}

  \def\Y#1S{\ensuremath{\PUpsilon{(#1S)}}\xspace}% no space before {...}!

\def\Lbar {\ensuremath{\kern 0.1em\overline{\kern -0.1em\PLambda}}\xspace}

\newcommand{\decay}[2]{\ensuremath{#1\!\to #2}\xspace}         % {\Pa}{\Pb \Pc}

\def\to                 {\ensuremath{\rightarrow}\xspace}

\def\eps   {\ensuremath{\varepsilon}\xspace}

\def\CP                {\ensuremath{C\!P}\xspace}
\def\CPT               {\ensuremath{C\!PT}\xspace}

\def\BdTopipi     {\decay{\Bd}{\pip\pim}}
\def\BdToKpi      {\decay{\Bd}{\Kp\pim}}
\def\BsToKK       {\decay{\Bs}{\Kp\Km}}

\def\AT#1     {\ensuremath{A_{\mathrm{T}}^{#1}}\xspace}           % 2

\def\C#1      {\ensuremath{\mathcal{C}_{#1}}\xspace}                       % 9
\def\Cp#1     {\ensuremath{\mathcal{C}_{#1}^{'}}\xspace}                    % 7
\def\Ceff#1   {\ensuremath{\mathcal{C}_{#1}^{\mathrm{(eff)}}}\xspace}        % 9  
\def\Cpeff#1  {\ensuremath{\mathcal{C}_{#1}^{'\mathrm{(eff)}}}\xspace}       % 7
\def\Ope#1    {\ensuremath{\mathcal{O}_{#1}}\xspace}                       % 2
\def\Opep#1   {\ensuremath{\mathcal{O}_{#1}^{'}}\xspace}                    % 7

\newcommand{\tev}{\ifthenelse{\boolean{inbibliography}}{\ensuremath{~T\kern -0.05em eV}\xspace}{\ensuremath{\mathrm{\,Te\kern -0.1em V}}\xspace}}
\newcommand{\gev}{\ensuremath{\mathrm{\,Ge\kern -0.1em V}}\xspace}
\newcommand{\mev}{\ensuremath{\mathrm{\,Me\kern -0.1em V}}\xspace}
\newcommand{\kev}{\ensuremath{\mathrm{\,ke\kern -0.1em V}}\xspace}
\newcommand{\ev}{\ensuremath{\mathrm{\,e\kern -0.1em V}}\xspace}
\newcommand{\gevc}{\ensuremath{{\mathrm{\,Ge\kern -0.1em V\!/}c}}\xspace}
\newcommand{\mevc}{\ensuremath{{\mathrm{\,Me\kern -0.1em V\!/}c}}\xspace}
\newcommand{\gevcc}{\ensuremath{{\mathrm{\,Ge\kern -0.1em V\!/}c^2}}\xspace}
\newcommand{\gevgevcccc}{\ensuremath{{\mathrm{\,Ge\kern -0.1em V^2\!/}c^4}}\xspace}
\newcommand{\mevcc}{\ensuremath{{\mathrm{\,Me\kern -0.1em V\!/}c^2}}\xspace}

\def\mum  {\ensuremath{\,\upmu\rm m}\xspace}

\def\invfb   {\ensuremath{\mbox{\,fb}^{-1}}\xspace}

\def\ps   {\ensuremath{{\rm \,ps}}\xspace}
\def\fs   {\ensuremath{\rm \,fs}\xspace}

\def\invps{\ensuremath{{\rm \,ps^{-1}}}\xspace}

\newcommand{\chisq}{\ensuremath{\chi^2}\xspace}

\newcommand{\chisqip}{\ensuremath{\chi^2_{\rm IP}}\xspace}

\def\gsim{{~\raise.15em\hbox{$>$}\kern-.85em
          \lower.35em\hbox{$\sim$}~}\xspace}
\def\lsim{{~\raise.15em\hbox{$<$}\kern-.85em
          \lower.35em\hbox{$\sim$}~}\xspace}

\def\evtgen     {\mbox{\textsc{EvtGen}}\xspace}

\def\geant      {\mbox{\textsc{Geant4}}\xspace}

\def\photos     {\mbox{\textsc{Photos}}\xspace}

\def\pythia     {\mbox{\textsc{Pythia}}\xspace}

\def\tell1  {TELL1\xspace}
\def\ukl1   {UKL1\xspace}

%% file: title-LHCb-PAPER.tex
% $Id: title-LHCb-PAPER.tex 35201 2013-05-10 14:33:25Z roldeman $
% ===============================================================================
% Purpose: LHCb-PAPER journal paper title page template
% Author: 
% Created on: 2010-09-25
% ===============================================================================

%%%%%%%%%%%%%%%%%%%%%%%%%
%%%%%  TITLE PAGE  %%%%%%
%%%%%%%%%%%%%%%%%%%%%%%%%
\begin{titlepage}
\pagenumbering{roman}

% Header ---------------------------------------------------
\vspace*{-1.5cm}
\centerline{\large EUROPEAN ORGANIZATION FOR NUCLEAR RESEARCH (CERN)}
\vspace*{1.5cm}
\hspace*{-0.5cm}
\begin{tabular*}{\linewidth}{lc@{\extracolsep{\fill}}r}
\ifthenelse{\boolean{pdflatex}}% Logo format choice
{\vspace*{-2.7cm}\mbox{\!\!\!\includegraphics[width=.14\textwidth]{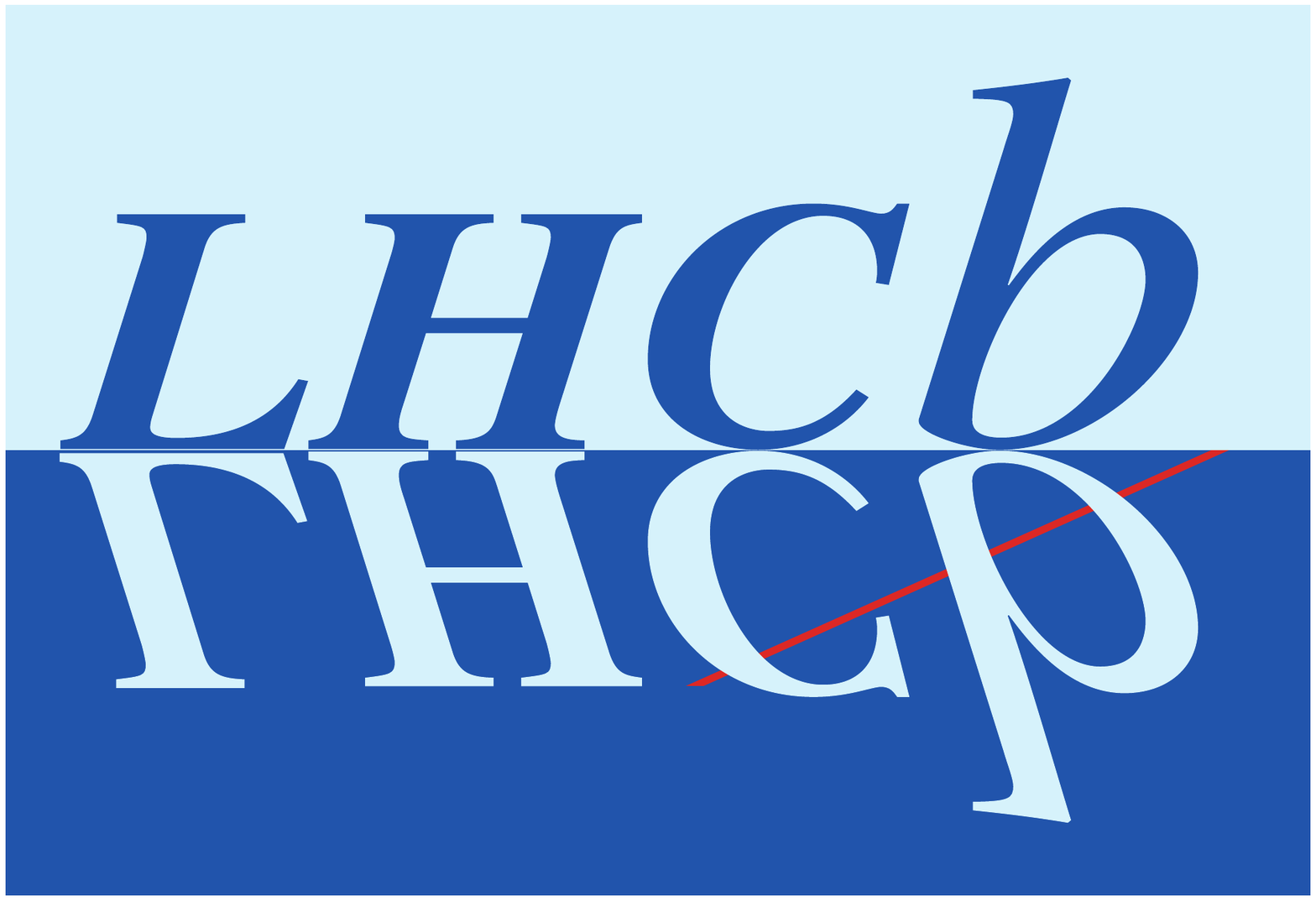}} & &}%
{\vspace*{-1.2cm}\mbox{\!\!\!\includegraphics[width=.12\textwidth]{figs/lhcb-logo.eps}} & &}%
\\
 & & CERN-PH-EP-2013-144 \\  % ID 
 & & LHCb-PAPER-2013-040 \\  % ID 
 & & 29 September 2013 \\ % Date - Can also hardwire e.g.: 23 March 2010
 & & \\
% not in paper \hline
\end{tabular*}

\vspace*{4.0cm}

% Title --------------------------------------------------
{\bf\boldmath\LARGE
\begin{center}
  First measurement of time-dependent\\$C\!P$ violation in $B^0_s \to K^+K^-$ decays
\end{center}
}

\vspace*{2.0cm}

% Authors -------------------------------------------------
\begin{center}
The LHCb collaboration\footnote{Authors are listed on the following pages.}
\end{center}

\vspace{\fill}

% Abstract -----------------------------------------------
\begin{abstract}
  \noindent
Direct and mixing-induced \CP-violating asymmetries in $B^0_s\to K^+K^-$ decays are measured for the first time using a data sample of $pp$ collisions, corresponding to an integrated luminosity of 1.0\invfb, collected with the LHCb detector at a centre-of-mass energy of 7\tev. The results are $C_{KK} = 0.14 \pm 0.11 \pm 0.03$ and $S_{KK} = 0.30 \pm 0.12 \pm 0.04$, where the first uncertainties are statistical and the second systematic. The corresponding quantities are also determined for $B^0\to \pi^+\pi^-$ decays to be $C_{\pi\pi} = -0.38 \pm 0.15 \pm 0.02$ and $S_{\pi\pi} = -0.71 \pm 0.13 \pm 0.02$, in good agreement with existing measurements.
\end{abstract}

\vspace*{2.0cm}

%\begin{center}
%  Submitted to JHEP
%\end{center}

\vspace{\fill}

{\footnotesize 
\centerline{\copyright~CERN on behalf of the \lhcb collaboration, license \href{http://creativecommons.org/licenses/by/3.0/}{CC-BY-3.0}.}}
\vspace*{2mm}

\end{titlepage}

%%%%%%%%%%%%%%%%%%%%%%%%%%%%%%%%
%%%%%  EOD OF TITLE PAGE  %%%%%%
%%%%%%%%%%%%%%%%%%%%%%%%%%%%%%%%

%  empty page follows the title page ----
\newpage
\setcounter{page}{2}
\mbox{~}
\newpage

% Author List ----------------------------
%  You need to get a new author list!
\input{LHCb_HD_authorlist_2013-06-20.tex}

\cleardoublepage

%% file: LHCb_HD_authorlist_2013-06-20.tex
%%%%%%%%%%%%%%%%%%%%%%%%%%%%%%%%%%%%%%%%%%
\centerline{\large\bf LHCb collaboration}
\begin{flushleft}
\small
R.~Aaij$^{40}$, 
B.~Adeva$^{36}$, 
M.~Adinolfi$^{45}$, 
C.~Adrover$^{6}$, 
A.~Affolder$^{51}$, 
Z.~Ajaltouni$^{5}$, 
J.~Albrecht$^{9}$, 
F.~Alessio$^{37}$, 
M.~Alexander$^{50}$, 
S.~Ali$^{40}$
% ...

{\footnotesize \it
% ...
$ ^{5}$Clermont Universit\'{e}, Universit\'{e} Blaise Pascal, CNRS/IN2P3, LPC, Clermont-Ferrand, France\\
$ ^{6}$CPPM, Aix-Marseille Universit\'{e}, CNRS/IN2P3, Marseille, France\\
% ...
$ ^{9}$Fakult\"{a}t Physik, Technische Universit\"{a}t Dortmund, Dortmund, Germany\\
% ...
$ ^{36}$Universidad de Santiago de Compostela, Santiago de Compostela, Spain\\
$ ^{37}$European Organization for Nuclear Research (CERN), Geneva, Switzerland\\
% ...
$ ^{40}$Nikhef National Institute for Subatomic Physics, Amsterdam, The Netherlands\\
% ...
$ ^{45}$H.H. Wills Physics Laboratory, University of Bristol, Bristol, United Kingdom\\
% ...
$ ^{50}$School of Physics and Astronomy, University of Glasgow, Glasgow, United Kingdom\\
$ ^{51}$Oliver Lodge Laboratory, University of Liverpool, Liverpool, United Kingdom\\
% ...
}
\end{flushleft}
%%%%%%%%%%%%%%%%%%%%%%%%%%%%%%%%%%%%%%%%%%

%% file: body.tex
\section{Introduction}
\label{sec:Introduction}

The study of $C\!P$ violation in charmless charged two-body decays of neutral $B$ mesons provides a test of the Cabibbo-Kobayashi-Maskawa (CKM) picture~\cite{Cabibbo:1963yz,*Kobayashi:1973fv} of the Standard Model (SM), and is a sensitive probe to contributions of processes beyond SM~\cite{Fleischer:1999pa,Gronau:2000md,Lipkin:2005pb,Fleischer:2007hj,Fleischer:2010ib}. However, quantitative SM predictions for \CP violation in these decays are challenging because of the presence of loop (penguin) amplitudes, in addition to tree amplitudes. As a consequence, the interpretation of the observables requires knowledge of hadronic factors that cannot be accurately calculated from quantum chromodynamics at present. Although this represents a limitation, penguin amplitudes may also receive contributions from non-SM physics. It is necessary to combine several measurements from such two-body decays, exploiting approximate flavour symmetries, in order to cancel or constrain the unknown hadronic factors~\cite{Fleischer:1999pa,Fleischer:2007hj}.

With the advent of the BaBar and Belle experiments, the isospin analysis of $B \to \pi\pi$ decays~\cite{Gronau:1990ka} has been one of the most important tools for determining the phase of the CKM matrix. As discussed in Refs.\cite{Fleischer:1999pa,Fleischer:2007hj,Fleischer:2010ib}, the hadronic parameters entering the $B^0 \to \pi^+\pi^-$ and $B^0_s \to K^+K^-$ decays are related by the U-spin symmetry, \emph{i.e.} by the exchange of $d$ and $s$ quarks in the decay diagrams. Although the U-spin symmetry is known to be broken to a larger extent than isospin, it is expected that the experimental knowledge of $B^0_s \to K^+K^-$ can improve the determination of the CKM phase, also in conjunction with the $B \to \pi\pi$ isospin analysis~\cite{Ciuchini:2012gd}.

Other precise measurements in this sector also provide valuable information for constraining hadronic parameters and give insights into hadron dynamics.
LHCb has already performed measurements of time-integrated \CP asymmetries in $B^0 \to K^+\pi^-$ and $B^0_s \to K^-\pi^+$ decays~\cite{LHCb-PAPER-2011-029,LHCb-PAPER-2013-018}, as well as measurements of branching fractions of charmless charged two-body $b$-hadron decays~\cite{LHCb-PAPER-2012-002}.

In this paper, the first measurement of time-dependent \CP-violating asymmetries in $B^0_s \to K^+K^-$ decays is presented. The analysis is based on a data sample, corresponding to an integrated luminosity of $1.0$\invfb, of $pp$ collisions at a centre-of-mass energy of 7\tev collected with the LHCb detector. A new measurement of the corresponding quantities for $B^0\to\pi^+\pi^-$ decays, previously measured with good precision by the BaBar~\cite{Lees:2013bb} and Belle~\cite{Adachi:2013mae} experiments, is also presented. The inclusion of charge-conjugate decay modes is implied throughout.

Assuming \CPT invariance, the \CP asymmetry as a function of time for neutral $B$ mesons decaying to a \CP eigenstate $f$ is given by
\begin{equation}
\mathcal{A}(t)=\frac{\Gamma_{{\Bb}^0_{(s)} \to f}(t)-\Gamma_{B^0_{(s)} \to f}(t)}{\Gamma_{{\Bb}^0_{(s)} \to f}(t)+\Gamma_{B^0_{(s)} \to f}(t)}=\frac{-C_f \cos(\Delta m_{d(s)} t) + S_f \sin(\Delta m_{d(s)} t)}{\cosh\left(\frac{\Delta\Gamma_{d(s)}}{2} t\right) - A^{\Delta\Gamma}_f \sinh\left(\frac{\Delta\Gamma_{d(s)}}{2} t\right)},
\end{equation}
where $\Delta m_{d(s)}=m_{{d(s)},\,\mathrm{H}}-m_{{d(s)},\,\mathrm{L}}$ and $\Delta\Gamma_{d(s)}=\Gamma_{{d(s)},\,\mathrm{L}}-\Gamma_{{d(s)},\,\mathrm{H}}$ are the mass and width differences of the $B^0_{(s)}$--$\Bb^0_{(s)}$ system mass eigenstates. The subscripts $\mathrm{H}$ and $\mathrm{L}$ denote the heaviest and lightest of these eigenstates, respectively. The quantities $C_f$, $S_f$ and $A^{\Delta\Gamma}_f$ are 
\begin{equation}
\begin{split}
C_{f} = \frac{1-|\lambda_f|^2}{1+|\lambda_f|^2},\,\,\,\,\,\,\,\,\,\,\,S_{f} =  \frac{2 {\rm Im} \lambda_f}{1+|\lambda_f|^2},\,\,\,\,\,\,\,\,\,\,\,A^{\Delta\Gamma}_f =  - \frac{2 {\rm Re} \lambda_f}{1+|\lambda_f|^2},
\end{split}\label{eq:adirmix} 
\end{equation}
with $\lambda_f$ defined as
\begin{equation}
\lambda_f = \frac{q}{p}\frac{\bar{A}_f}{A_f}.
\end{equation}
The two mass eigenstates of the effective Hamiltonian in the $B^0_{(s)}$--$\Bb^0_{(s)}$ system are $p|B^0_{(s)}\rangle \pm q|\Bb^0_{(s)}\rangle$, where $p$ and $q$ are complex parameters. The parameter $\lambda_f$ is thus related to $B^0_{(s)}$--$\Bb^0_{(s)}$ mixing (via $q/p$) and to the decay amplitudes of the $B^0_{(s)} \to f$ decay ($A_f$) and of the $\Bb^0_{(s)} \to f$ decay ($\bar{A}_f$). Assuming, in addition, negligible $C\!P$ violation in the mixing ($|q/p|=1$), as expected in the SM and confirmed by current experimental determinations~\cite{bib:hfagbase,Aaij:2013gta}, the terms $C_{f}$ and $S_{f}$ parameterize direct and mixing-induced \CP violation, respectively. In the case of the $B^0_s \to K^+K^-$ decay, these terms can be expressed as~\cite{Fleischer:1999pa}
\begin{equation}
C_{KK}=\frac{2\tilde{d}^{\prime}\sin\vartheta^{\prime}\sin\gamma}{1+2\tilde{d}^{\prime}\cos\vartheta^{\prime}\cos\gamma+\tilde{d}^{\prime2}},\label{eq:adirkksm}
\end{equation}
\begin{equation}
S_{KK}=\frac{\sin(2\beta_{s}-2\gamma)+2\tilde{d}^{\prime}\cos\vartheta^{\prime}\sin(2\beta_{s}-\gamma)+\tilde{d}^{\prime2}\sin(2\beta_{s})}
{1+2\tilde{d}^{\prime}\cos\vartheta^{\prime}\cos\gamma+\tilde{d}^{\prime2}},\label{eq:amixkksm}\end{equation}
where $\tilde{d}^{\prime}$ and $\vartheta^{\prime}$ are hadronic parameters related to the magnitude and phase of the tree and penguin amplitudes, respectively, $-2\beta_{s}$ is the $B^0_s$--$\Bb^0_s$ mixing phase, and $\gamma$ is the angle of the unitarity triangle given by $\arg \left[ - \left(V_{ud}V_{ub}^*\right) /  \left(V_{cd}V_{cb}^*\right) \right ]$.
Additional information can be provided by the knowledge of $A^{\Delta\Gamma}_{KK}$, determined from $B^0_s \to K^+K^-$ effective lifetime measurements~\cite{LHCb-PAPER-2011-014,LHCb-PAPER-2012-013}.

The paper is organized as follows. After a brief introduction on the detector, trigger and simulation in Sec.~\ref{sec:dettrigev}, the event selection is described in Sec.~\ref{sec:selection}. The measurement of time-dependent \CP asymmetries with neutral $B$ mesons requires that the flavour of the decaying $B$ meson at the time of production is identified. This is discussed in Sec.~\ref{sec:flavourTagging}. Direct and mixing-induced \CP asymmetry terms are determined by means of two maximum likelihood fits to the invariant mass and decay time distributions: one fit for the $B^0_s \to K^+K^-$ decay and one for $B^0 \to \pi^+\pi^-$ decay. The fit model is described in Sec.~\ref{sec:fitModel}. In Sec.~\ref{sec:tagcalib}, the calibration of flavour tagging performances, realized by using a fit to $B^0 \to K^+\pi^-$  and $B^0_s \to K^-\pi^+$ mass and decay time distributions, is discussed. The results of the \BsToKK and  \BdTopipi fits are given in Sec.~\ref{sec:fitresults} and the determination of systematic uncertainties discussed in Sec.~\ref{sec:systematics}. Finally, conclusions are drawn in Sec.~\ref{sec:conclusions}.

\section{Detector, trigger and simulation}
\label{sec:dettrigev}
The \lhcb detector~\cite{Alves:2008zz} is a single-arm forward spectrometer covering the \mbox{pseudorapidity} range between 2 and 5,
designed for the study of particles containing \bquark or \cquark quarks. The detector includes a high-precision tracking system
consisting of a silicon-strip vertex detector surrounding the $pp$ interaction region, a large-area silicon-strip detector located
upstream of a dipole magnet with a bending power of about $4{\rm\,Tm}$, and three stations of silicon-strip detectors and straw
drift tubes placed downstream.

The combined tracking system provides a momentum measurement with relative uncertainty that varies from 0.4\% at 5\gevc to 0.6\% at 100\gevc,
and impact parameter ($d_\mathrm{IP}$) resolution of 20\mum for tracks with high transverse momenta. The $d_\mathrm{IP}$ is defined as the minimum distance between the reconstructed trajectory of a particle and a given $pp$ collision vertex~(PV). Charged hadrons are identified
using two ring-imaging Cherenkov (RICH) detectors~\cite{LHCb-DP-2012-003}. Photon, electron and hadron candidates are identified by a calorimeter system consisting of
scintillating-pad and preshower detectors, an electromagnetic calorimeter and a hadronic calorimeter. Muons are identified by a system composed of alternating layers of iron and multiwire
proportional chambers~\cite{LHCb-DP-2012-002}.

The trigger~\cite{LHCb-DP-2012-004} consists of a hardware stage, based on information from the calorimeter and muon systems, followed by a software stage, which applies a full event reconstruction. Events selected by any hardware trigger decision are included in the analysis. The software trigger requires a two-, three- or four-track secondary vertex with a large sum of the transverse momenta of the tracks and a significant displacement from the PVs.
At least one track should have a transverse momentum ($p_\mathrm{T}$) exceeding $1.7$\gevc and \chisqip with respect to any PV greater than 16. The \chisqip is the
difference in \chisq of a given PV reconstructed with and without the considered track.

A multivariate algorithm~\cite{BBDT} is used for the identification of secondary vertices consistent with the decay of a \bquark hadron. To improve the trigger efficiency on hadronic two-body $B$ decays, a dedicated two-body software trigger is also used.
This trigger selection imposes requirements on the following quantities: the quality of the reconstructed tracks (in terms of $\chi^2$/ndf, where ndf is the number of degrees of freedom), their $p_\mathrm{T}$ and $d_\mathrm{IP}$; the distance of closest approach of the decay products of the $B$ meson candidate ($d_\mathrm{CA}$), its transverse momentum ($p_\mathrm{T}^B$), impact parameter ($d_\mathrm{IP}^B$) and the decay time in its rest frame ($t_{\pi\pi}$, calculated assuming decay into $\pi^+\pi^-$). 

Simulated events are used to determine the signal selection efficiency as a function of the decay time, and to study flavour tagging, decay time resolution and background modelling. In the simulation, $pp$ collisions are generated using \pythia~6.4~\cite{Sjostrand:2006za} with a specific \lhcb configuration~\cite{LHCb-PROC-2010-056}. Decays of hadronic particles
are described by \evtgen~\cite{Lange:2001uf}, in which final state radiation is generated using \photos~\cite{Golonka:2005pn}. The interaction of the generated particles with the detector and its response are implemented using the \geant toolkit~\cite{Allison:2006ve, *Agostinelli:2002hh} as described in Ref.~\cite{LHCb-PROC-2011-006}.

\section{Event selection}
\label{sec:selection}

Events passing the trigger requirements are filtered to reduce the size of the data sample by means of a loose preselection. 
Candidates that pass the preselection are then classified into mutually exclusive samples of different final states by means of the particle identification (PID) capabilities of the \rich detectors. Finally, a boosted decision tree~(BDT) algorithm~\cite{Breiman} is used to separate signal from background.

Three sources of background are considered: other two-body $b$-hadron decays with a misidentified pion or kaon in the final state (cross-feed background), pairs of randomly associated oppositely-charged tracks (combinatorial background), and pairs of oppositely-charged tracks from partially reconstructed three-body $B$ decays (three-body background). Since the three-body background gives rise to candidates with invariant mass values well separated from the signal mass peak, the event selection is mainly intended to reject cross-feed and combinatorial backgrounds, which mostly affect the invariant mass region around the nominal $B^0_{(s)}$ mass.

The preselection, in addition to tighter requirements on the kinematic variables already used in the software trigger, applies requirements on the largest $p_\mathrm{T}$ and on the largest $d_\mathrm{IP}$ of the $B$ candidate decay products, as summarized in Table~\ref{tab:strippingCuts}.

\begin{table}
\begin{centering}
\caption{\small Kinematic requirements applied by the event preselection.}
  \begin{tabular}{c|c}
   Variable & Requirement \\
    \hline
Track $\chi^2$/ndf & $<5$ \\
     Track $p_\mathrm{T}\,[\!\gevc]$ & $>1.1$ \\
   Track $d_\mathrm{IP}\,[\!\mum]$ & $>120$ \\
   $\mathrm{max}\,p_\mathrm{T}\,[\!\gevc]$ & $>2.5$ \\
   $\mathrm{max}\,d_\mathrm{IP}\,[\!\mum]$ & $>200$ \\
   $d_\mathrm{CA}\,[\!\mum]$ & $<80$ \\
    $p_\mathrm{T}^{B}\,\,[\!\gevc]$  & $>1.2$ \\
   $d_\mathrm{IP}^{B}\,[\!\mum]$ & $<100$ \\
    $t_{\pi\pi}\,\,[\textrm{ps}]$  & $>0.6$ \\
   $m_{\pi^+\pi^-}\,\,[\!\gevcc]$ & $4.8$--$5.8$ \\
  \end{tabular}
  \par\end{centering}
\label{tab:strippingCuts}
\end{table}

\begin{figure}[t]
  \begin{center}
    \includegraphics[width=0.49\textwidth]{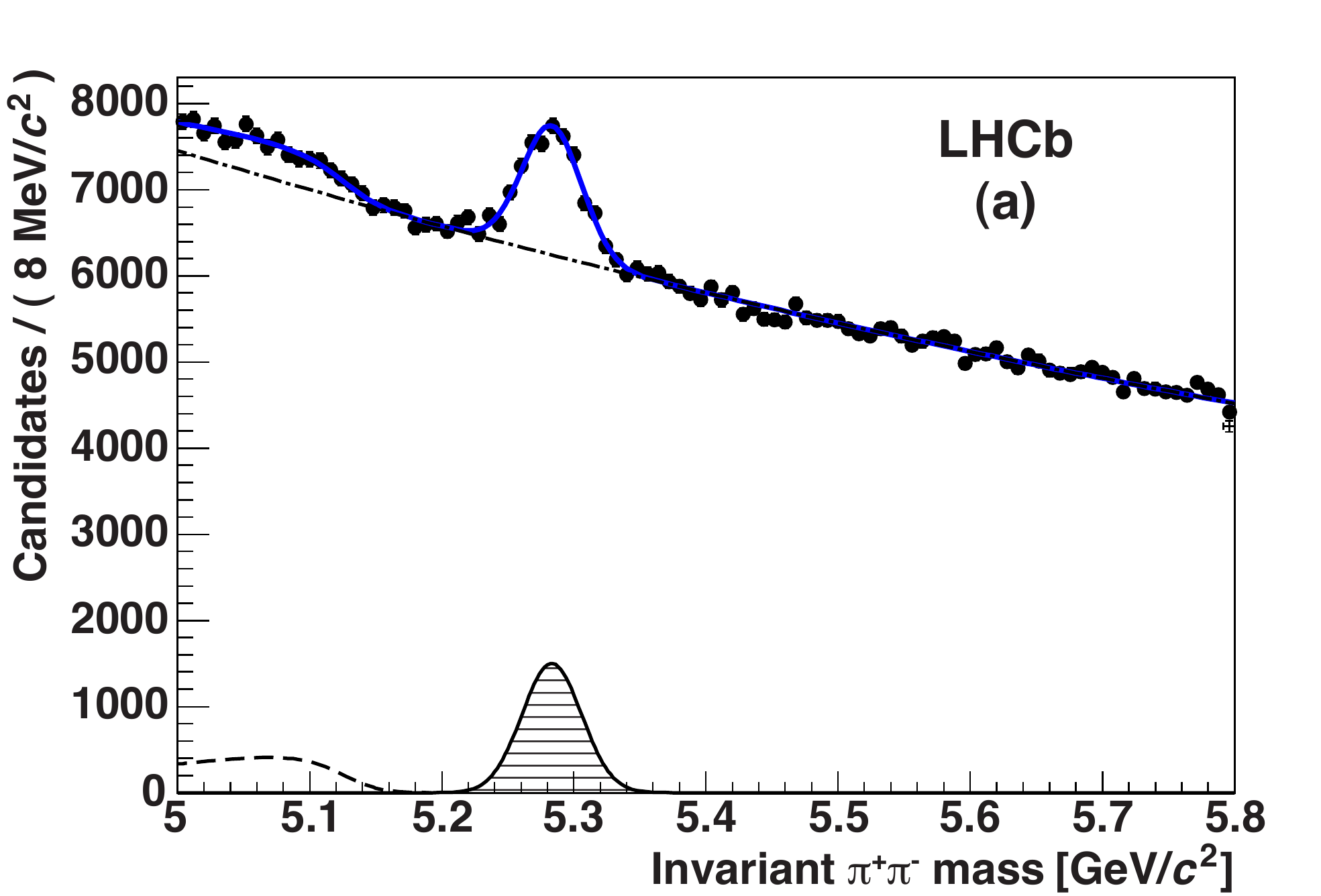}
    \includegraphics[width=0.49\textwidth]{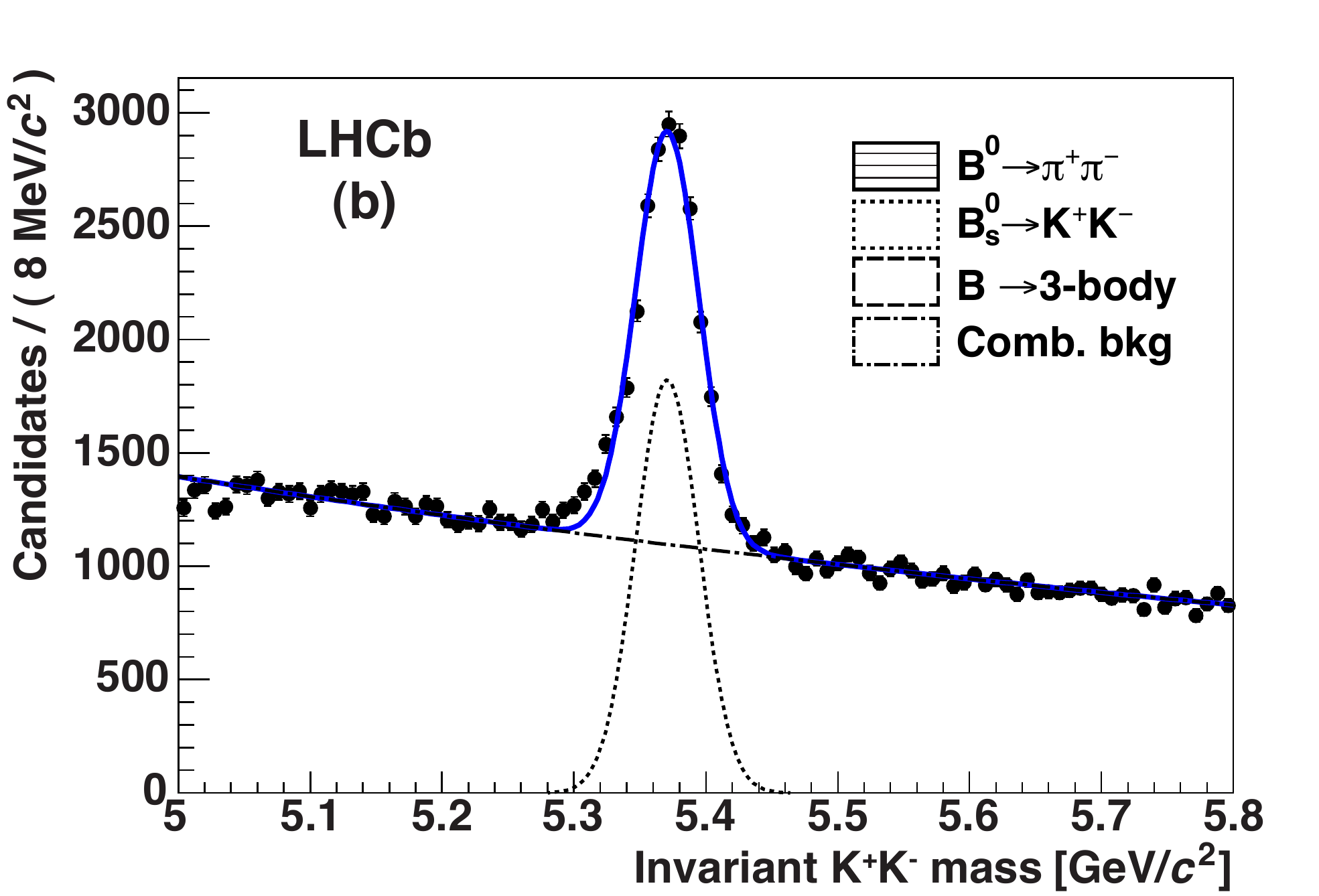}
 \end{center}
  \caption{\small Fits to the (a) $\pi^+\pi^-$ and (b) $K^+K^-$ invariant mass spectra, after applying preselection and PID requirements. The components contributing to the fit model are shown.}
  \label{fig:normalization}
\end{figure}

The main source of cross-feed background in the \BdTopipi and \BsToKK invariant mass signal regions is the \BdToKpi decay, where one of the two final state particles is misidentified. The PID is able to reduce this background to 15\% (11\%) of the $B^0_s \to K^+K^-$ ($B^0 \to \pi^+\pi^-$) signal. Invariant mass fits are used to estimate the yields of signal and combinatorial components. Figure~\ref{fig:normalization} shows the $\pi^+\pi^-$ and $K^+K^-$ invariant mass spectra after applying preselection and PID requirements. The results of the fits, which use a single Gaussian function to describe the signal components and neglect residual backgrounds from cross-feed decays, are superimposed. The presence of a small component due to partially reconstructed three-body decays in the $K^+K^-$ spectrum is also neglected. Approximately $11\times 10^3$ $B^0 \to \pi^+\pi^-$ and $14\times 10^3$ $B^0_s \to K^+K^-$ decays are reconstructed.

A BDT discriminant based on the AdaBoost algorithm~\cite{AdaBoost} is then used to reduce the combinatorial background. The BDT uses the following properties of the decay products: the minimum $p_\mathrm{T}$ of the pair, the minimum $d_\mathrm{IP}$, the minimum \chisqip, the maximum $p_\mathrm{T}$, the maximum $d_\mathrm{IP}$, the maximum \chisqip, the $d_\mathrm{CA}$, and the $\chi^2$ of the common vertex fit. The BDT also uses the following properties of the $B$ candidate: the $p_\mathrm{T}^B$, the $d_\mathrm{IP}^B$, the \chisqip, the flight distance, and the $\chi^2$ of the flight distance. The BDT is trained, separately for the \BdTopipi and the \BsToKK decays, using simulated events to model the signal and data in the mass sideband ($5.5 < m < 5.8$\gevcc) to model the combinatorial background.
An optimal threshold on the BDT response is then chosen by maximizing $S/\sqrt{S+B}$, where $S$ and $B$ represent the numbers of signal and combinatorial background events within $\pm 60$\mevcc (corresponding to about $\pm 3\sigma$) around the \Bd or \Bs mass.
The resulting mass distributions are discussed in Sec.~\ref{sec:fitresults}. A control sample of $B^0 \to K^+\pi^-$ and $B^0_s \to K^-\pi^+$ decays is selected using the BDT selection optimized for the $B^0 \to \pi^+\pi^-$ decay, but with different PID requirements applied.

\section{Flavour tagging}\label{sec:flavourTagging}

The sensitivity to the time-dependent \CP asymmetry is directly related to the tagging power, defined as $\eps_{\rm eff} = \eps(1-2\omega)^2$, where $\eps$ is the probability that a tagging decision for a given candidate can be made (tagging efficiency) and $\omega$ is the probability that such a decision is wrong (mistag probability). If the candidates are divided into different subsamples, each one characterized by an average tagging efficiency $\eps_i$ and an average mistag probability $\omega_i$, the effective tagging power is given by $\eps_{\rm eff} = \sum_i \eps_i(1-2\omega_i)^2$, where the index $i$ runs over the various subsamples.

So-called opposite-side (OS) taggers are used to determine the initial flavour of the signal $B$ meson~\cite{Aaij:2012mu}. This is achieved by looking at the charge of the lepton, either muon or electron, originating from semileptonic decays, and of the kaon from the $b \rightarrow c \rightarrow s$ decay transition
of the other $b$ hadron in the event. An additional OS tagger, the vertex
charge tagger, is based on the inclusive reconstruction of the opposite $B$
decay vertex and on the computation of a weighted average of the charges of all tracks associated
to that vertex. For each tagger, the mistag probability is estimated by means of an artificial neural network. When more than one tagger is available per candidate, 
these probabilities are combined into a single mistag probability $\eta$ and a unique decision per candidate is taken.

The data sample is divided into tagging categories according to the value of $\eta$, and a calibration is performed to obtain the corrected mistag probability $\omega$ for each category by means of a mass and decay time fit to the $B^0 \to K^+\pi^-$ and $B^0_s \to K^-\pi^+$ spectra, as described in Sec.~\ref{sec:tagcalib}. The consistency of tagging performances for \BdTopipi, \BsToKK, \BdToKpi and $B^0_s \to K^-\pi^+$ decays is verified using simulation. The definition of tagging categories is optimized to obtain the highest tagging power. This is achieved by the five categories reported in Table~\ref{tab:categoryDefinition}. The gain in tagging power using more categories is found to be marginal. 

\begin{table}[t]
  \caption{\small Definition of the five tagging categories determined from the optimization algorithm, in terms of ranges of the mistag probability $\eta$.}
  \begin{center}
    \begin{tabular}{c|c}
      Category & Range for $\eta$ \\
      \hline
      1 & $0.00 - 0.22 $ \\
      2 & $0.22 - 0.30 $ \\
      3 & $0.30 - 0.37 $ \\
      4 & $0.37 - 0.42 $ \\
      5 & $0.42 - 0.47 $ 
    \end{tabular}
  \end{center}
  \label{tab:categoryDefinition}
\end{table}

\section{Fit model}\label{sec:fitModel}

For each component that contributes to the selected samples, the distributions of invariant mass, decay time and tagging decision are modelled.
Three sources of background are considered: combinatorial background, cross-feed and backgrounds from partially reconstructed three-body decays. The following cross-feed backgrounds play a non-negligible role:
\begin{itemize}
\item in the $K^\pm\pi^\mp$ spectrum, \BdTopipi and \BsToKK decays where one of the two final state particles is misidentified, and $B^0 \to K^+\pi^-$ decays where pion and kaon identities are swapped;
\item in the $\pi^+\pi^-$ spectrum, \BdToKpi decays where the kaon is misidentified as a pion; in this spectrum there is also a small component of $B^0_s \to \pi^+\pi^-$ which must be taken into account~\cite{LHCb-PAPER-2012-002};
\item in the $K^+K^-$ spectrum, \BdToKpi decays where the pion is misidentified as a kaon.
\end{itemize}

\subsection{Mass model}

The signal component for each two-body decay is modelled convolving a double Gaussian function with a parameterization of final state QED radiation. The probability density function (PDF) is given by
\begin{equation}
g(m)=A\left [\Theta(\mu-m)\,\left (\mu-m \right )^s \right] \otimes G_2(m;\,f_1,\,\sigma_1,\,\sigma_2),\label{eq:radcor}
\end{equation}
where $A$ is a normalization factor, $\Theta$ is the Heaviside function, $G_2$ is the sum of two Gaussian functions with widths $\sigma_1$ and $\sigma_2$ and zero mean, $f_1$ is the fraction of the first Gaussian function, and $\mu$ is the $B$-meson mass. The negative parameter $s$ governs the amount of final state QED radiation, and is fixed for each signal component using the respective theoretical QED prediction, calculated according to Ref.~\cite{Baracchini:2005wp}.

The combinatorial background is modelled by an exponential function for all the final states. The component due to partially reconstructed three-body \B
decays in the $\pi^+\pi^-$ and $K^+K^-$ spectra is modelled convolving a Gaussian resolution function with an ARGUS function~\cite{Albrecht:1989ga}. The $K^\pm\pi^\mp$ spectrum is described convolving a Gaussian function with the sum of two ARGUS functions, in order to accurately model not only \Bd and $B^+$, but also a smaller fraction of \Bs three-body decays~\cite{LHCb-PAPER-2013-018}. Cross-feed background PDFs are obtained from simulations. For each final state hypothesis, a set of invariant mass distributions is determined from pairs where one or both tracks are misidentified, and each of them is parameterized by means of a kernel estimation technique~\cite{Cranmer:2000du}.
The yields of the cross-feed backgrounds are fixed by means of a time-integrated simultaneous fit to the mass spectra of all two-body $B$ decays~\cite{LHCb-PAPER-2013-018}.

\subsection{Decay time model}

The time-dependent decay rate of a flavour-specific $B\to f$ decay and of its \CP conjugate ${\Bb}\to\bar{f}$ is given by the PDF
\begin{equation}\label{eq:decayTimeB2KPI}
  \begin{split}
    f\left(t,\,\psi,\,\xi\right) & = K\left(1-\psi A_{\CP}\right)\left(1-\psi A_{f}\right) \times \\
    & \left\lbrace\! \left[\left(1\!-\!A_\mathrm{P}\right)\!\Omega_{\xi}^{B}\!+\!\left(1\!+\!A_\mathrm{P}\right)\!\bar{\Omega}_{\xi}^{B}\right]\!H_{+}\left(t\right)\!+\!
      \psi \!\left[\left(1\!-\!A_\mathrm{P}\right)\!\Omega_{\xi}^{B}\!-\!\left(1\!+\!A_\mathrm{P}\right)\!\bar{\Omega}_{\xi}^{B}\right]\!H_{-}\left(t\right) \!\right \rbrace, 
  \end{split}
\end{equation}
where $K$ is a normalization factor, and the variables $\psi$ and $\xi$ are the final state tag and the initial flavour tag, respectively. This PDF is suitable for the cases of $B^0 \to K^+\pi^-$ and $B^0_s \to K^-\pi^+$ decays. The variable $\psi$ assumes the value $+1$ for the final state $f$ and $-1$ for the final state $\bar{f}$. The variable $\xi$ assumes the discrete value $+k$ when the candidate is tagged as \B in the $k$-th category, $-k$ when the candidate is tagged as \Bb in the $k$-th category, and zero for untagged candidates. The direct \CP asymmetry, $A_{\CP}$, the asymmetry of final state reconstruction efficiencies (detection asymmetry), $A_{f}$, and the \B meson production asymmetry, $A_\mathrm{P}$, are defined as
\begin{eqnarray}
  A_{\CP} & = & \frac{\mathcal{B}\left({\Bb}\to\bar{f}\right)-\mathcal{B}\left(B\to f\right)}{\mathcal{B}\left({\Bb}\to\bar{f}\right)+\mathcal{B}\left(B\to f\right)}, \\
  A_{f} & = & \frac{\varepsilon_\mathrm{rec}\left(\bar{f}\right)-\varepsilon_\mathrm{rec}\left(f\right)}{\varepsilon_\mathrm{rec}\left(\bar{f}\right)+\varepsilon_\mathrm{rec}\left(f\right)}, \\
  A_\mathrm{P} & = & \frac{\mathcal{R}\left({\Bb}\right)-\mathcal{R}\left(B\right)}{\mathcal{R}\left({\Bb}\right)+\mathcal{R}\left(B\right)},
\end{eqnarray}
where $\mathcal{B}$ denotes the branching fraction, $\varepsilon_\mathrm{rec}$ is the reconstruction efficiency of the final state $f$ or $\bar{f}$, and $\mathcal{R}$ is the production rate of the given \B or ${\Bb}$ meson.
The parameters $\Omega_{\xi}^{B}$ and $\bar{\Omega}_{\xi}^{B}$ are the probabilities that a \B or a \Bb meson is tagged as $\xi$, respectively, and are defined as
\begin{equation}
  \begin{split}
\Omega_{k}^{B} = \varepsilon_{k}\left(1-\omega_{k}\right),\qquad \Omega_{-k}^{B} = \varepsilon_{k}\omega_{k},\qquad \Omega_{0}^{B} = 1-\sum_{j=1}^5\varepsilon_{j},\\
\bar{\Omega}_{k}^{B} = \bar{\varepsilon}_{k}\bar{\omega}_{k},\qquad \bar{\Omega}_{-k}^{B} = \bar{\varepsilon}_{k}\left(1-\bar{\omega}_{k}\right),\qquad \bar{\Omega}_{0}^{B} = 1-\sum_{j=1}^5\bar{\varepsilon}_{j},
  \end{split}
\end{equation}
where $\varepsilon_{k}$ ($\bar{\varepsilon}_{k}$) is the tagging efficiency and $\omega_{k}$ ($\bar{\omega}_{k}$) is the mistag probability for signal \B (\Bb) mesons that belong to the $k$-th tagging category.
The functions $H_{+}\left(t\right)$ and $H_{-}\left(t\right)$ are defined as
\begin{eqnarray}
  H_{+}\left(t\right) & = & \left[e^{-\Gamma_{d(s)} t}\cosh{\left(\Delta\Gamma_{d(s)}t/2\right)}\right]\otimes R\left(t\right) \varepsilon_\mathrm{acc}\left(t\right), \\
  H_{-}\left(t\right) & = & \left[e^{-\Gamma_{d(s)} t}\cos{\left(\Delta m_{d(s)}t\right)}\right]\otimes R\left(t\right) \varepsilon_\mathrm{acc}\left(t\right),
\end{eqnarray}
where $\Gamma_{d(s)}$ is the average decay width of the $B^0_{(s)}$ meson, $R$ is the decay time resolution model, and $\varepsilon_\mathrm{acc}$ is the decay time acceptance. 

In the fit to the $B^0 \to K^+\pi^-$ and $B^0_s \to K^-\pi^+$ mass and decay time distributions, the decay width differences of \Bd and \Bs mesons are fixed to zero and to the value measured by \lhcb, $\Delta\Gamma_{s} = 0.106$\invps~\cite{LHCb-PAPER-2013-002}, respectively, whereas the mass differences are left free to vary. The fit is performed simultaneously for candidates belonging to the five tagging categories and for untagged candidates.

If the final states $f$ and $\bar{f}$ are the same, as in the cases of $B^0 \to \pi^+\pi^-$ and $B^0_s \to K^+K^-$ decays, the time-dependent decay rates are described by
\begin{equation}\label{eq:decayTimeeigen}
  f\left(t,\,\xi\right) = K\left\lbrace\left[\left(1\!-\!A_\mathrm{P}\right)\!\Omega_{\xi}^{B}\!+\!\left(1\!+\!A_\mathrm{P}\right)\!\bar{\Omega}_{\xi}^{B}\right]\!I_{+}\left(t\right)\!+\!\left[\left(1\!-\!A_\mathrm{P}\right)\!\Omega_{\xi}^{B}\!-\!\left(1\!+\!A_\mathrm{P}\right)\!\bar{\Omega}_{\xi}^{B}\right]\!I_{-}\left(t\right)\right\rbrace,
\end{equation}
where the functions $I_{+}\left(t\right)$ and $I_{-}\left(t\right)$ are
\begin{eqnarray}
  I_{+}\left(t\right) \!&\! =\! &\! \left\{e^{-\Gamma_{d(s)} t}\!\left[\cosh\!{\left( \Delta\Gamma_{d(s)}t/2 \right)}\!-\!A_{f}^{\Delta\Gamma}\sinh\!{\left( \Delta\Gamma_{d(s)}t/2 \right)}\right]\!\right\}\!\otimes\! R\left(t\right) \varepsilon_\mathrm{acc}\left(t\right), \\
  I_{-}\left(t\right) \!&\! =\! &\! \left\{e^{-\Gamma_{d(s)} t}\!\left[C_f\cos\!{\left(\Delta m_{d(s)}t\right)}\!-\!S_f\sin\!{\left(\Delta m_{d(s)}t\right)}\right]\!\right\}\!\otimes\! R\left(t\right)\varepsilon_\mathrm{acc}\left(t\right).
\end{eqnarray}

In the $B^0_s \to K^+K^-$ fit, the average decay width and mass difference of the \Bs meson are fixed to the values $\Gamma_{s} = 0.661$\invps~\cite{LHCb-PAPER-2013-002} and $\Delta m_{s} = 17.768$\invps~\cite{Aaij:2013mpa}. The width difference $\Delta\Gamma_s$ is left free to vary, but is constrained to be positive as expected in the SM and measured by LHCb~\cite{XIE}, in order to resolve the invariance of the decay rates under the exchange $\left(\Delta\Gamma_{d(s)},\,A_{f}^{\Delta\Gamma}\right)\rightarrow\left(-\Delta\Gamma_{d(s)},\,-A_{f}^{\Delta\Gamma}\right)$. 
Moreover, the definitions of $C_f$, $S_f$ and $A_{f}^{\Delta\Gamma}$ in Eq.~(\ref{eq:adirmix}) allow $A_{f}^{\Delta\Gamma}$ to be expressed as
\begin{equation}
  A_{f}^{\Delta\Gamma} = \pm\sqrt{1-C_f^{2}-S_f^{2}}.
\end{equation}
The positive solution, which is consistent with measurements of the $B^0_s \to K^+K^-$ effective lifetime~\cite{LHCb-PAPER-2011-014,LHCb-PAPER-2012-013}, is taken.
In the case of the $B^0 \to \pi^+\pi^-$ decay, where the width difference of the \Bd meson is negligible and fixed to zero, the ambiguity is not relevant. The mass difference is fixed to the value $\Delta m_{d} = 0.516$\invps~\cite{Aaij:2012nt}. Again, these two fits are performed simultaneously for candidates belonging to the five tagging categories and for untagged candidates.

The dependence of the reconstruction efficiency on the decay time (decay time acceptance) is studied with simulated events. For each simulated decay, namely \BdTopipi, \BsToKK, \BdToKpi and $B^0_s \to K^- \pi^+$, reconstruction, trigger requirements and event selection are applied, as for data.
It is empirically found that $\varepsilon_\mathrm{acc}\left(t\right)$ is well parameterized by
\begin{equation}\label{eq:acceptance}
  \varepsilon_\mathrm{acc}\left(t\right) = 
  \frac{1}{2}\left[1-\frac{1}{2}\mathrm{erf}\left(\frac{p_{1}-t}{p_{2}\,t}\right)-\frac{1}{2}\mathrm{erf}\left(\frac{p_{3}-t}{p_{4}\,t}\right)\right],
\end{equation}
where $\mathrm{erf}$ is the error function, and $p_i$ are free parameters determined from simulation.

The expressions for the decay time PDFs of the cross-feed background components are determined from Eqs.~(\ref{eq:decayTimeB2KPI}) and~(\ref{eq:decayTimeeigen}), assuming that the decay time calculated under the wrong mass hypothesis resembles the correct one with sufficient accuracy. This assumption is verified with simulations.

The parameterization of the decay time distribution for combinatorial background events is studied using the high-mass sideband from data, defined as $5.5 < m < 5.8$\gevcc. Concerning the $K^\pm\pi^\mp$ spectrum, for events selected by the \BdTopipi BDT, it is empirically found that the PDF can be written as
\begin{equation}\label{eq:combinatorialTime}
  f\left(t,\,\xi, \,\psi\right) = K \Omega_{\xi}^\mathrm{comb}\left ( 1-\psi A_{CP}^\mathrm{comb} \right ) \left [ g\, e^{-\Gamma_{1}^\mathrm{comb}t}+\left(1-g\right)e^{-\Gamma_{2}^\mathrm{comb}t} \right ]\varepsilon^\mathrm{comb}_\mathrm{acc}(t),
\end{equation}
where $A_{\CP}^\mathrm{comb}$ is the charge asymmetry of the combinatorial background, $g$ is the fraction of the first exponential component, and $\Gamma_1^\mathrm{comb}$ and $\Gamma_2^\mathrm{comb}$ are two free parameters.
The term $\Omega_{\xi}^\mathrm{comb}$ is the probability to tag a background event as $\xi$. It is parameterized as
\begin{equation}
  \Omega_{k}^\mathrm{comb} = \varepsilon_{k}^\mathrm{comb},\qquad \Omega_{-k}^\mathrm{comb} = \bar{\varepsilon}_{k}^\mathrm{comb},\qquad \Omega_{0}^\mathrm{comb} = 1-\sum_{j}^5{\left ( \varepsilon_{j}^\mathrm{comb}+\bar{\varepsilon}_{j}^\mathrm{comb} \right )},\label{eq:omegabkg}
\end{equation}
where $\varepsilon_{k}^\mathrm{comb}$ ($\bar{\varepsilon}_{k}^\mathrm{comb}$) is the probability to tag a background event as a \B (\Bb) in the $k$-th category. 
The effective function $\varepsilon^\mathrm{comb}_\mathrm{acc}\left(t\right)$ is the analogue of the decay time acceptance for signal decays, and is given by the same expression of Eq.~(\ref{eq:acceptance}), but characterized by independent values of the parameters $p_i$.
For the $\pi^+\pi^-$ and $K^+K^-$ spectra, the same expression as in Eq.~(\ref{eq:combinatorialTime}) is used, with the difference that the charge asymmetry is set to zero and no dependence on $\psi$ is needed. 
   
The last case to examine is that of the three-body partially reconstructed backgrounds in the $K^\pm\pi^\mp$, $\pi^+\pi^-$, and $K^+K^-$ spectra. 
In the $K^\pm\pi^\mp$ mass spectrum there are two components, each described by an ARGUS function~\cite{Albrecht:1989ga}. Each of the two corresponding decay time components is empirically parameterized as
\begin{equation}\label{eq:physTime}
  f\left(t,\,\xi,\,\psi\right) = K \Omega_{\xi}^\mathrm{part} \left ( 1-\psi A_{\CP}^\mathrm{part} \right ) e^{-\Gamma^\mathrm{part}t}\varepsilon_\mathrm{acc}^\mathrm{part}\left(t\right),
\end{equation}
where $A_{\CP}^\mathrm{part}$ is the charge asymmetry and $\Gamma^\mathrm{part}$ is a free parameter.
The term $\Omega_{\xi}^\mathrm{part}$ is the probability to tag a background event as $\xi$, and is parameterized as in Eq.~(\ref{eq:omegabkg}), but with independent tagging probabilities.
For the $\pi^+\pi^-$ and $K^+K^-$ spectra, the same expression as in Eq.~(\ref{eq:physTime}) is used, with the difference that the charge asymmetry is set to zero and no dependence on $\psi$ is needed. 

\begin{figure}[t]
  \begin{center}
    \includegraphics[width=0.3\textwidth]{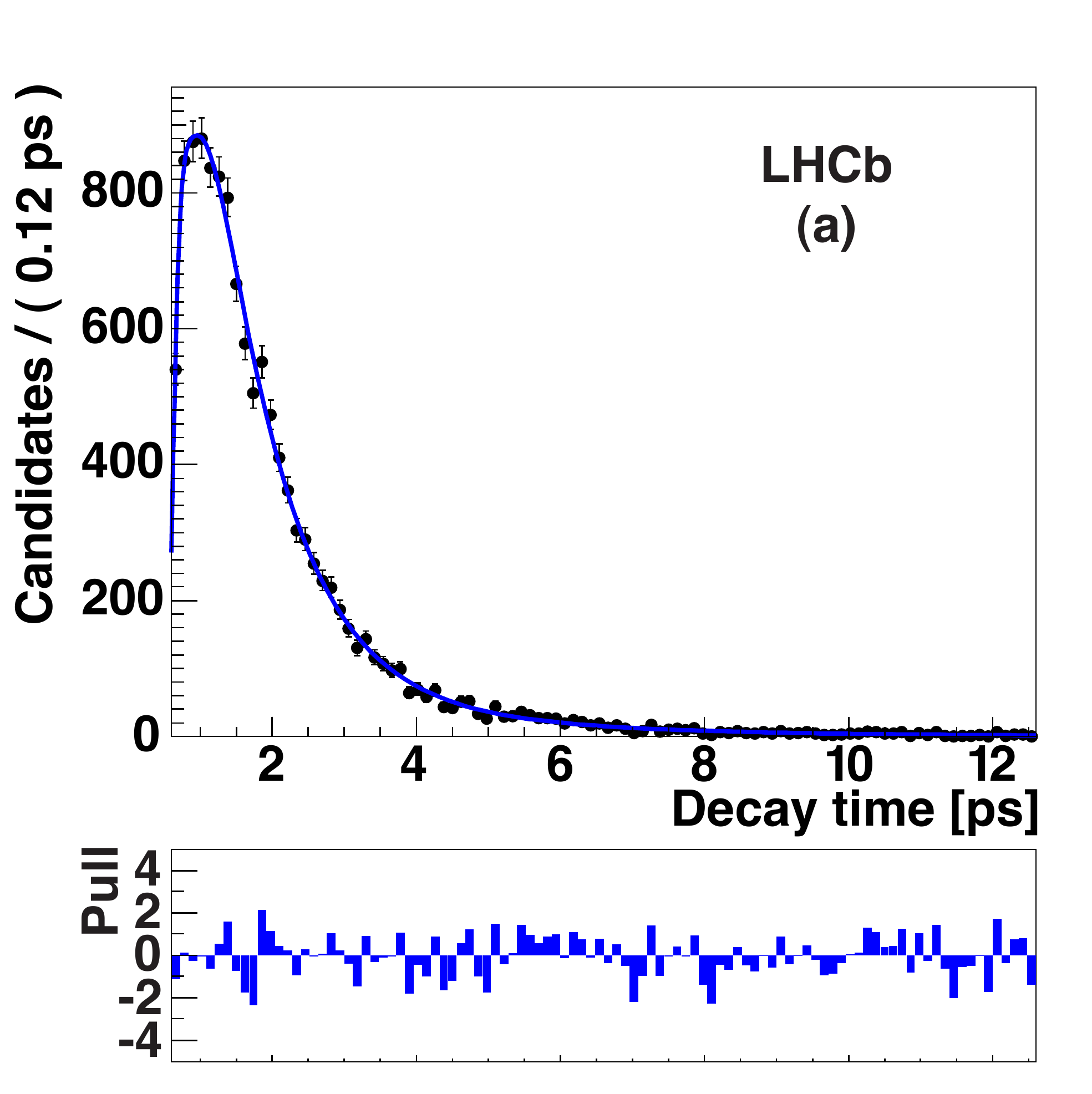}
    \includegraphics[width=0.3\textwidth]{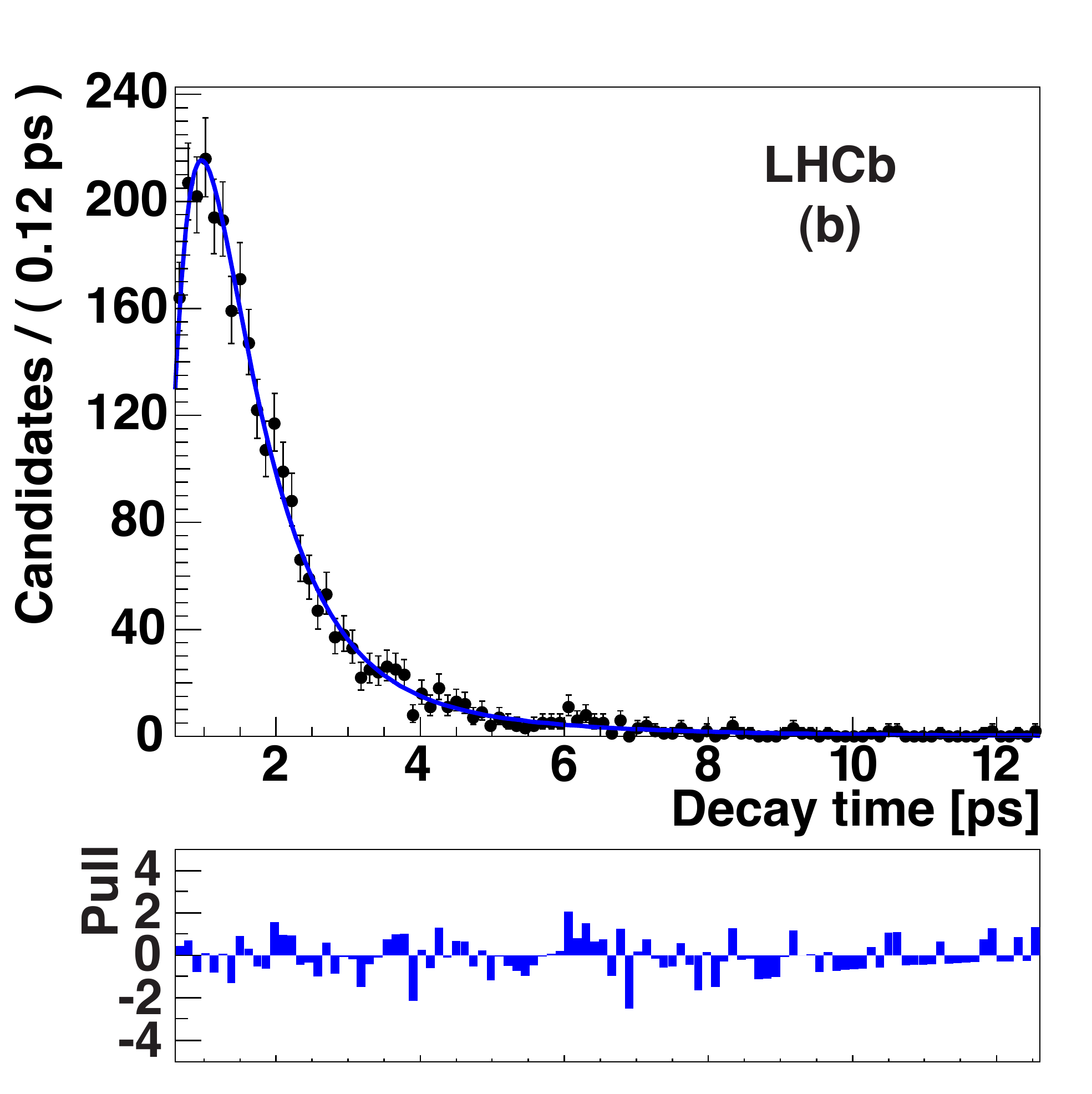}
    \includegraphics[width=0.3\textwidth]{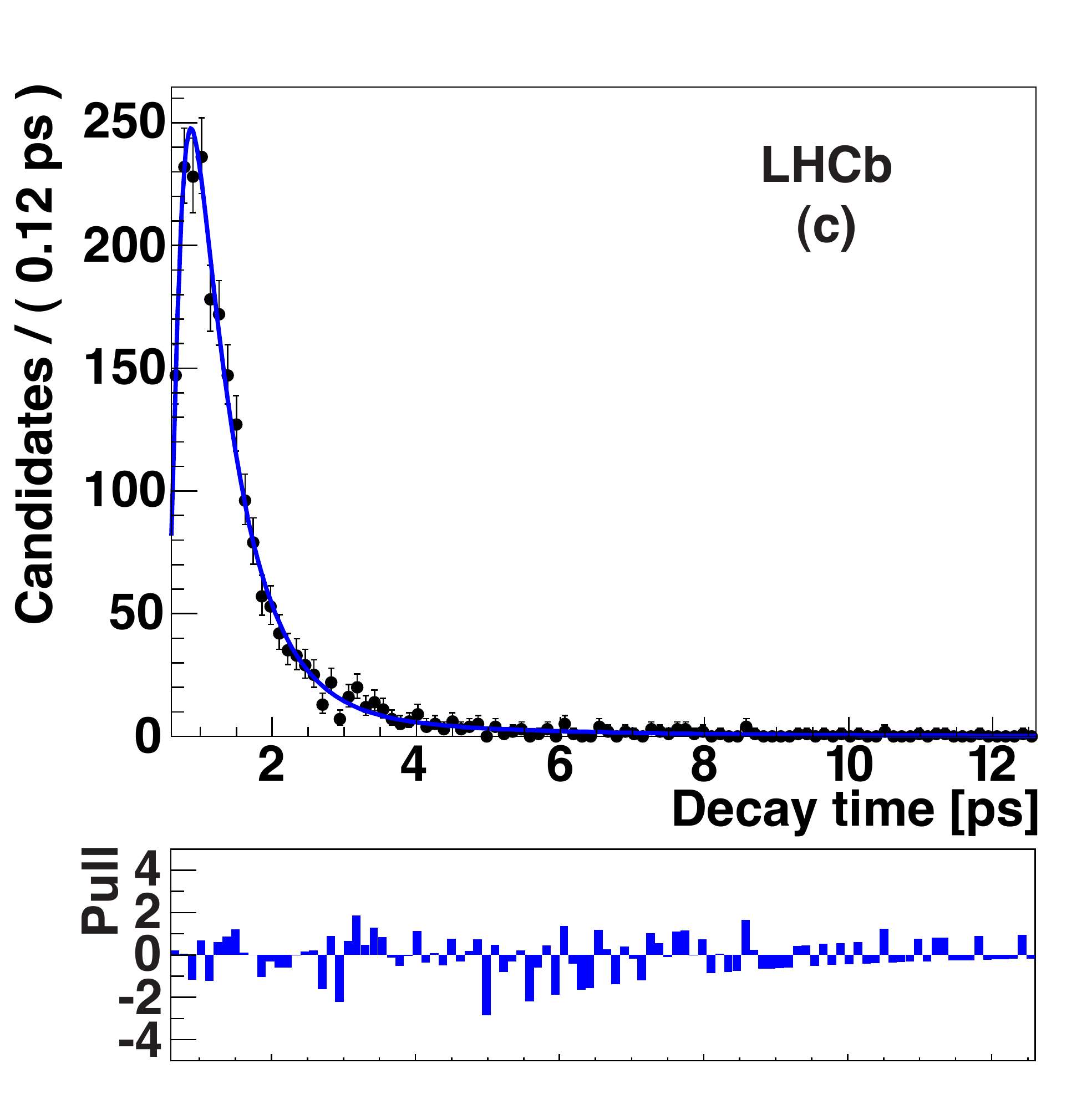}
   \includegraphics[width=0.3\textwidth]{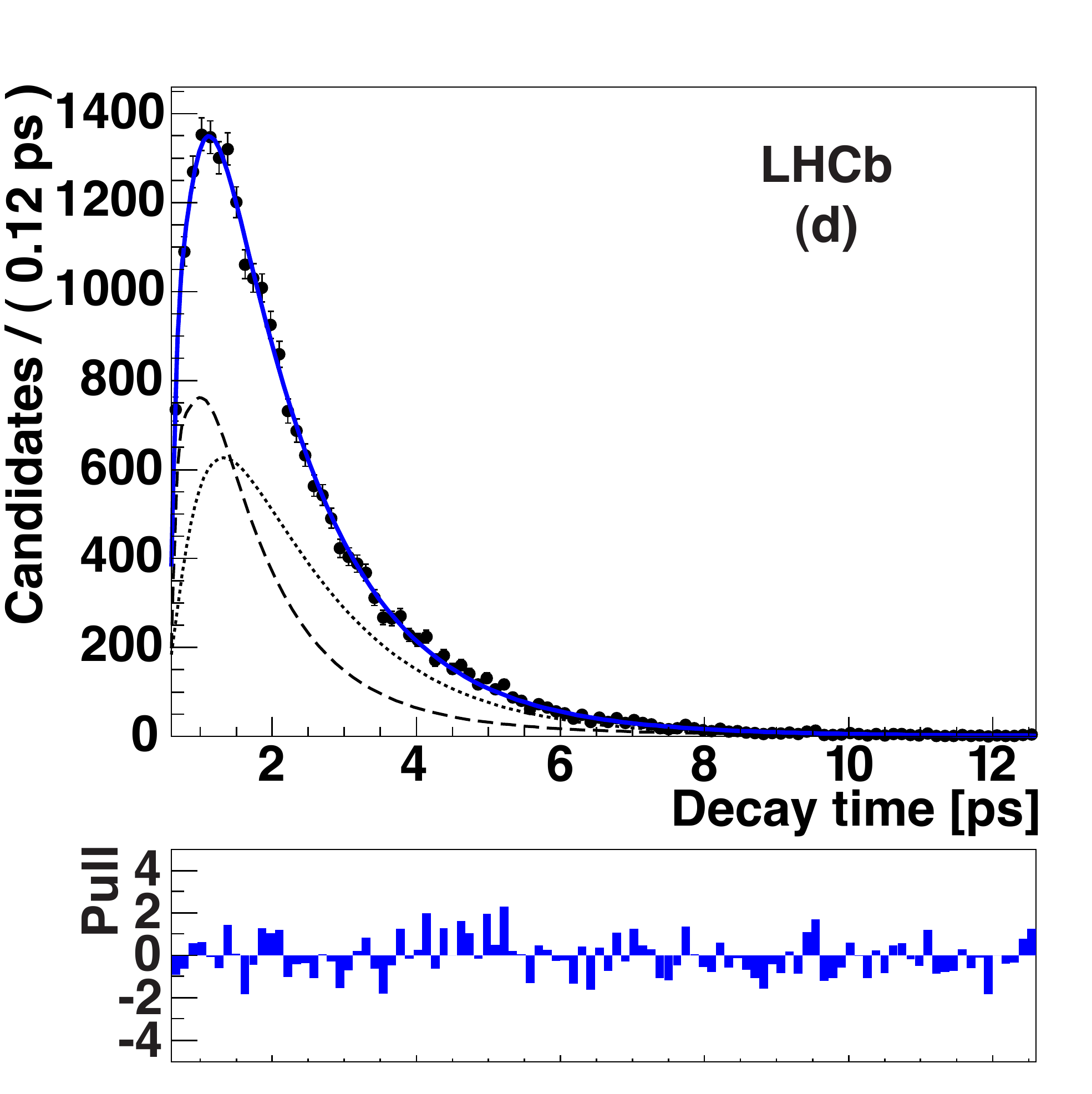}
   \includegraphics[width=0.3\textwidth]{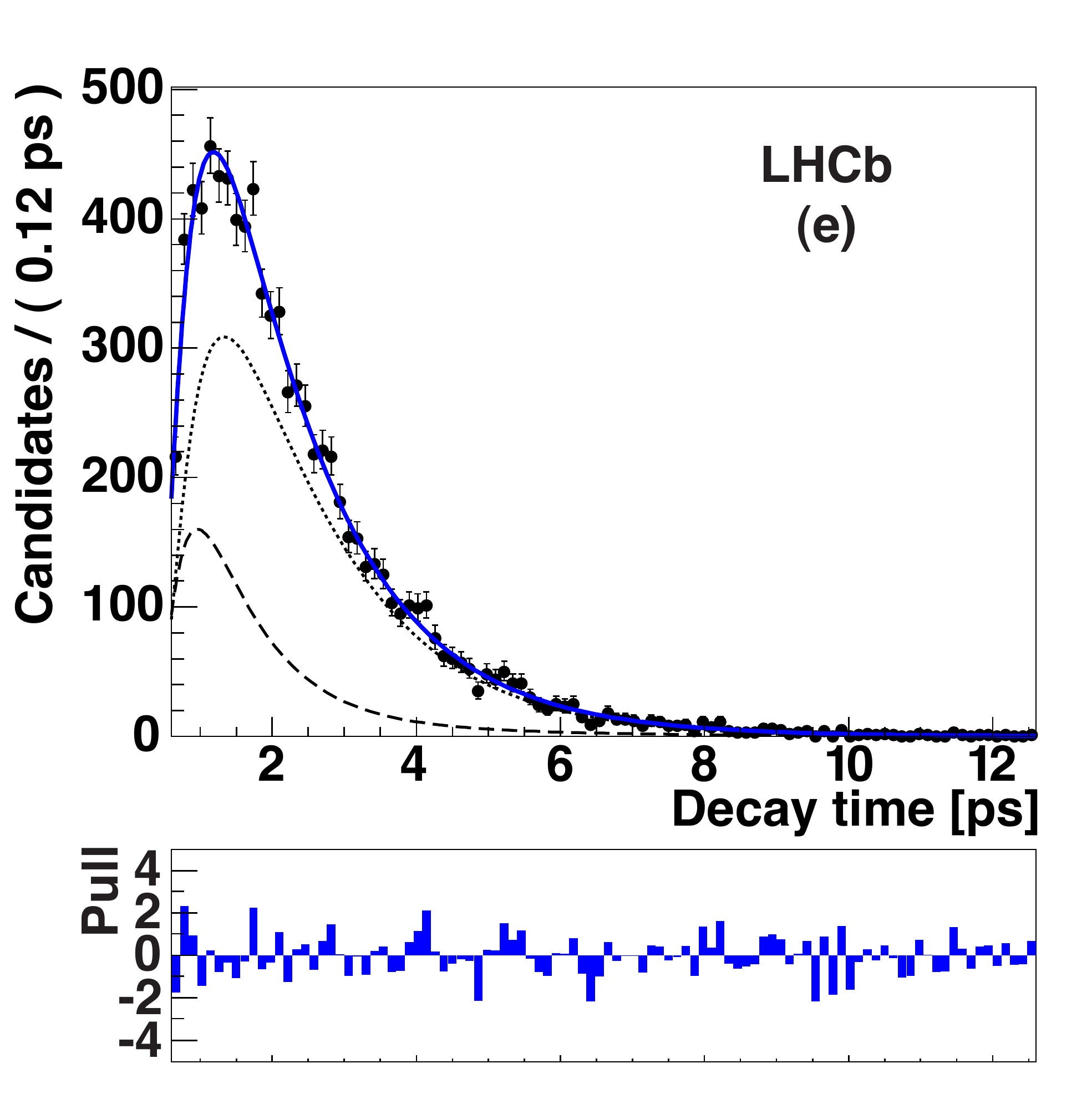}
    \includegraphics[width=0.3\textwidth]{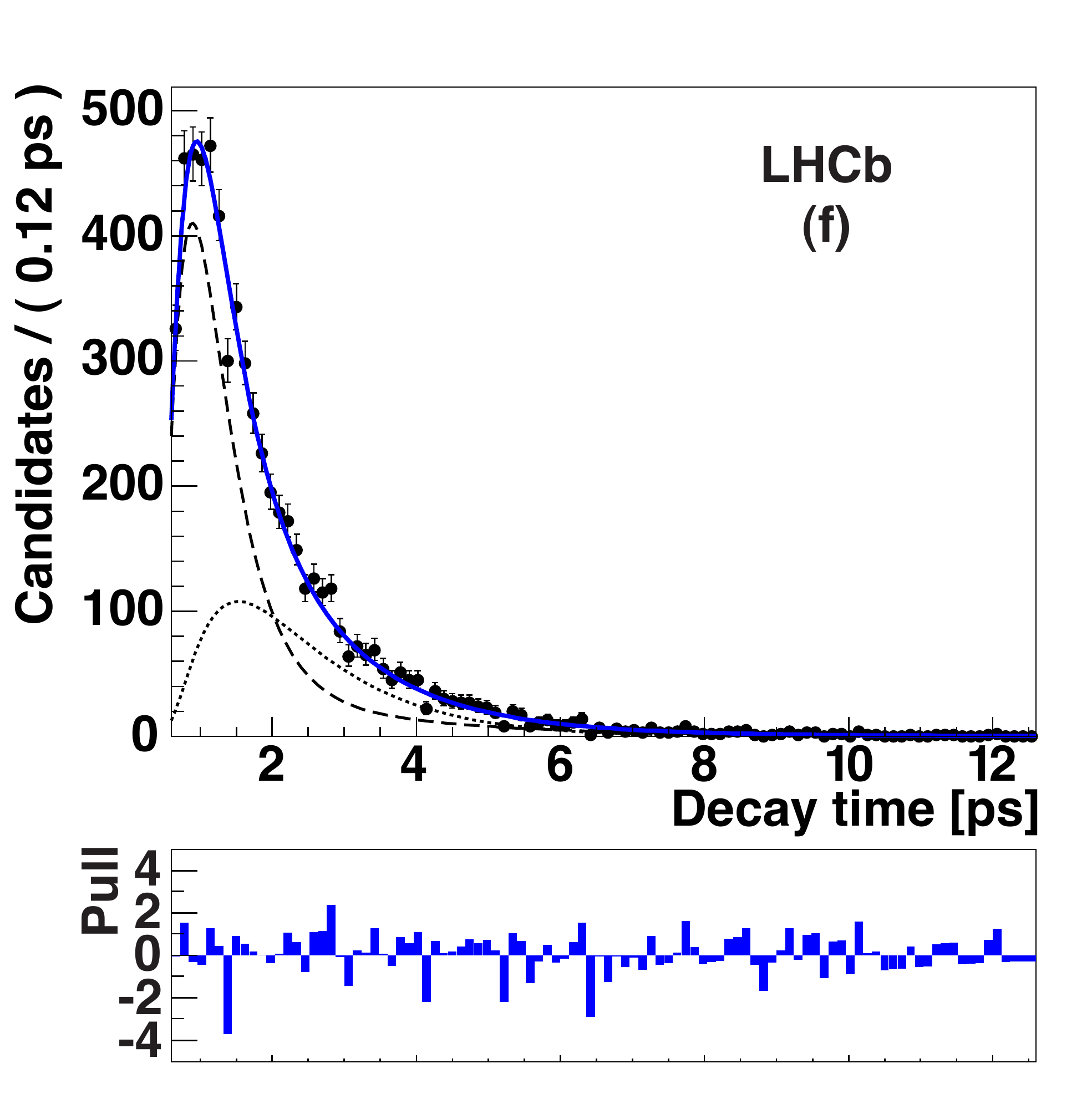}
 \end{center}
  \caption{\small Decay time distributions corresponding to (a, b, c) high- and (d, e, f) low-mass sidebands from the (a and d) $K^\pm\pi^\mp$, (b and e) $\pi^+\pi^-$ and (c and f) $K^+K^-$ mass spectra, with the results of fits superimposed. In the bottom plots, the combinatorial background component (dashed) and the three-body background component (dotted) are shown.}
  \label{fig:combinatorialandphysModel}
\end{figure}

The accuracy of the combinatorial and three-body decay time parameterizations is checked by performing a simultaneous fit to the invariant mass and decay time spectra of the high- and low-mass sidebands. The combinatorial background contributes to both the high- and low-mass sidebands, whereas the three-body background is only present in the low-mass side. In Fig.~\ref{fig:combinatorialandphysModel} the decay time distributions are shown, restricted to the high and low $K^\pm\pi^\mp$, $\pi^+\pi^-$, and $K^+K^-$ mass sidebands. The low-mass sidebands are defined by the requirement $5.00< m<5.15$\gevcc for $K^\pm\pi^\mp$ and $\pi^+\pi^-$, and by the requirement $5.00 < m<5.25$\gevcc for $K^+K^-$, whereas in all cases the high-mass sideband is defined by the requirement $5.5 < m < 5.8$\gevcc.

\subsection{Decay time resolution}
\label{sec:decaytimeres}

Large samples of $J/\psi\to \mu^+\mu^-$, $\psi(\mathrm{2S})\to \mu^+\mu^-$, $\Upsilon(\mathrm{1S})\to\mu^+\mu^-$, $\Upsilon(\mathrm{2S})\to\mu^+\mu^-$ and $\Upsilon(\mathrm{3S})\to\mu^+\mu^-$ decays can be selected without any requirement that biases the decay time. 
Maximum likelihood fits to the invariant mass and decay time distributions allow an average resolution to be derived for each of these decays.
A comparison of the resolutions determined from data and simulation yields correction factors ranging from $1.0$ to $1.1$, depending on the charmonium or bottomonium decay considered. On this basis, a correction factor $1.05 \pm 0.05$ is estimated. 
The simulation also indicates that, in the case of $B^0\to\pi^+\pi^-$ and $B_s^0\to K^+ K^-$ decays, an additional dependence of the resolution on the decay time must be considered. Taking this dependence into account, we finally estimate a decay time resolution of $50 \pm 10$\fs. Furthermore, from the same fits to the charmonium and bottomium decay time spectra, it is found that the measurement of the decay time is biased by less than $2$\fs. As a baseline resolution model, $R(t)$, a single Gaussian function with zero mean and $50$\fs  width is used. Systematic uncertainties on the direct and mixing-induced \CP-violating asymmetries in $B^0_s \to K^+K^-$ and $B^0 \to \pi^+\pi^-$ decays, related to the choice of the baseline resolution model, are discussed in Sec.~\ref{sec:systematics}.

\section{Calibration of flavour tagging}
\label{sec:tagcalib}

In order to measure time-dependent \CP asymmetries in \BdTopipi and \BsToKK decays, simultaneous unbinned maximum likelihood fits to the invariant mass and decay time distributions are performed. First, a fit to the $K^\pm\pi^\mp$ mass and time spectra is performed to determine the performance of the flavour tagging and the $B^0$ and $B^0_s$ production asymmetries. The flavour tagging efficiencies, mistag probabilities and production asymmetries are then propagated to the \BdTopipi and \BsToKK fits by multiplying the likelihood functions with Gaussian terms. The flavour tagging variables are parameterized as
\begin{equation}
\begin{split}
  \varepsilon_{k} = \varepsilon_{k}^\mathrm{tot}\left(1-A_{k}^{\varepsilon}\right),\qquad \bar{\varepsilon}_{k} = \varepsilon_{k}^\mathrm{tot}\left(1+A_{k}^{\varepsilon}\right), \\
  \omega_{k} = \omega_{k}^\mathrm{tot}\left(1-A_{k}^{\omega}\right),\qquad \bar{\omega}_{k} = \omega_{k}^\mathrm{tot}\left(1+A_{k}^{\omega}\right),
\end{split}
\end{equation} 
where $\varepsilon_{k}^\mathrm{tot}$ ($\omega_{k}^\mathrm{tot}$) is the tagging efficiency (mistag fraction) averaged between $B^0_{(s)}$ and $\Bb^0_{(s)}$ in the $k$-th category, and $A_{k}^{\varepsilon}$ ($A_{k}^{\omega}$) measures a possible asymmetry between the tagging efficiencies (mistag fractions) of $B^0_{(s)}$ and $\Bb^0_{(s)}$ in the $k$-th category.

\begin{figure}[t]
  \begin{center}
    \includegraphics[width=0.49\textwidth]{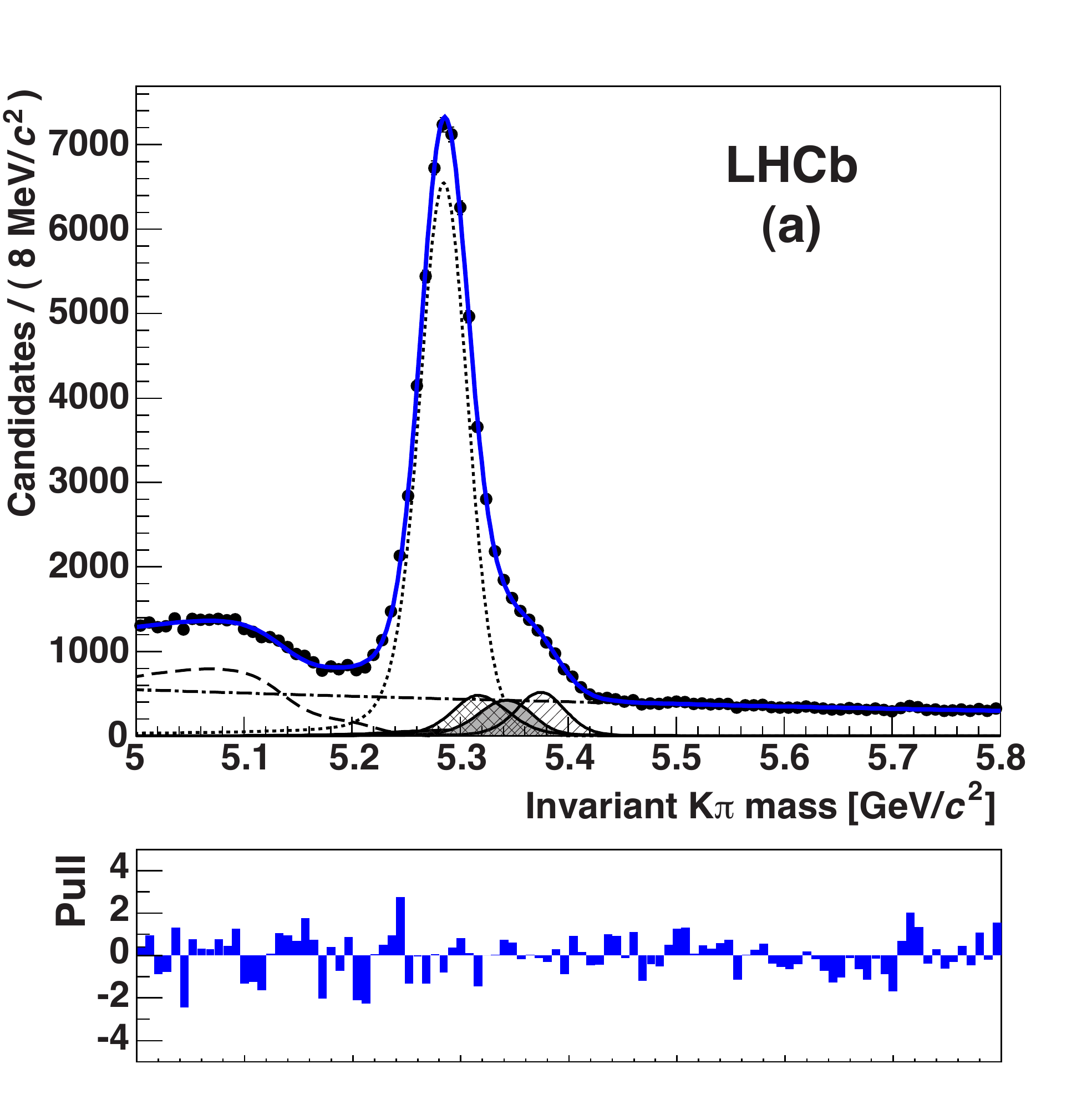}
   \includegraphics[width=0.49\textwidth]{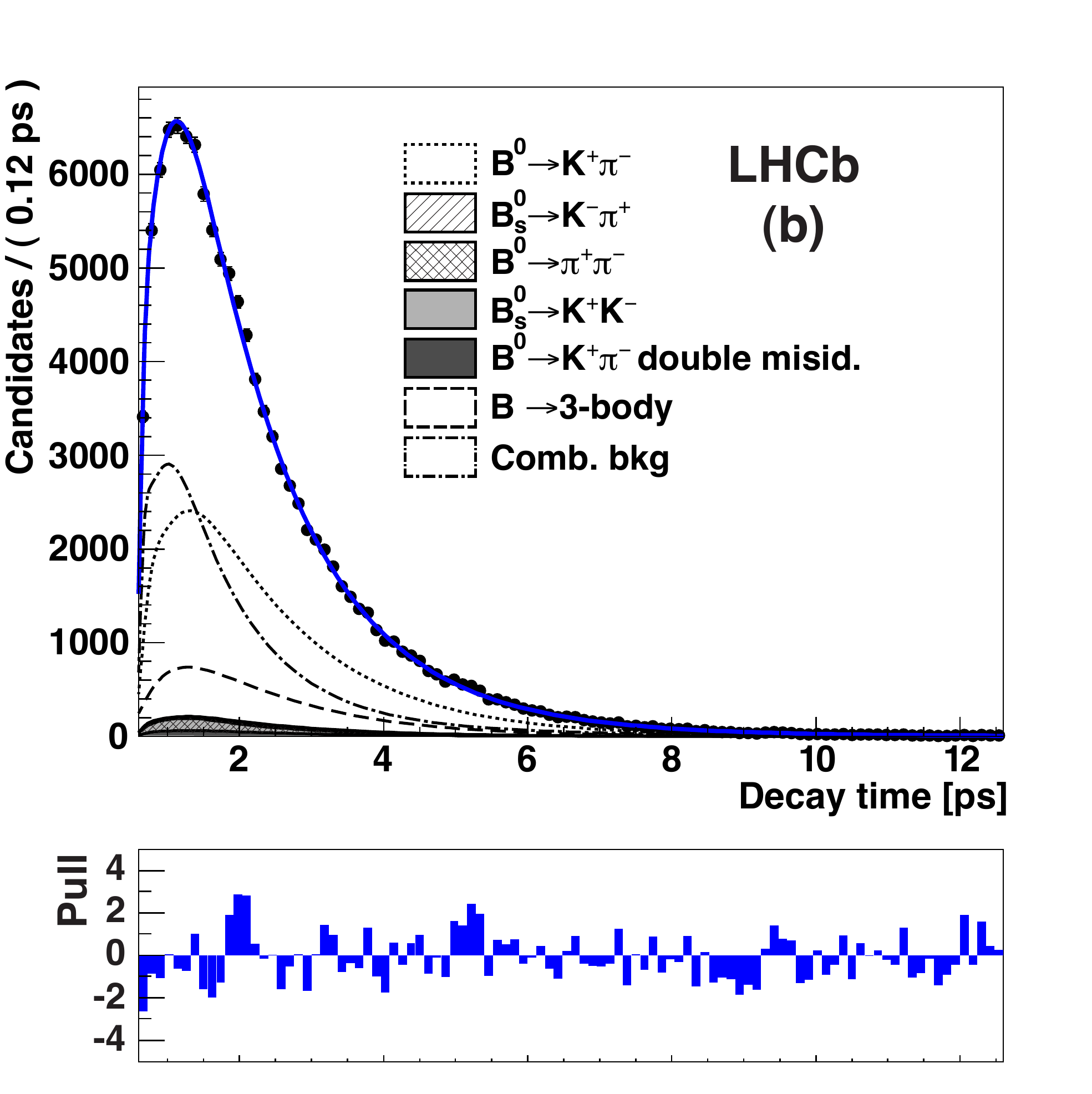}
 \end{center}
  \caption{\small Distributions of $K^\pm\pi^\mp$  (a) mass and (b) decay time, with the result of the fit overlaid. The main components contributing to the fit model are also shown.}
  \label{fig:bdkpiFit}
\end{figure}

\begin{figure}[t]
  \begin{center}
   \includegraphics[width=0.45\textwidth]{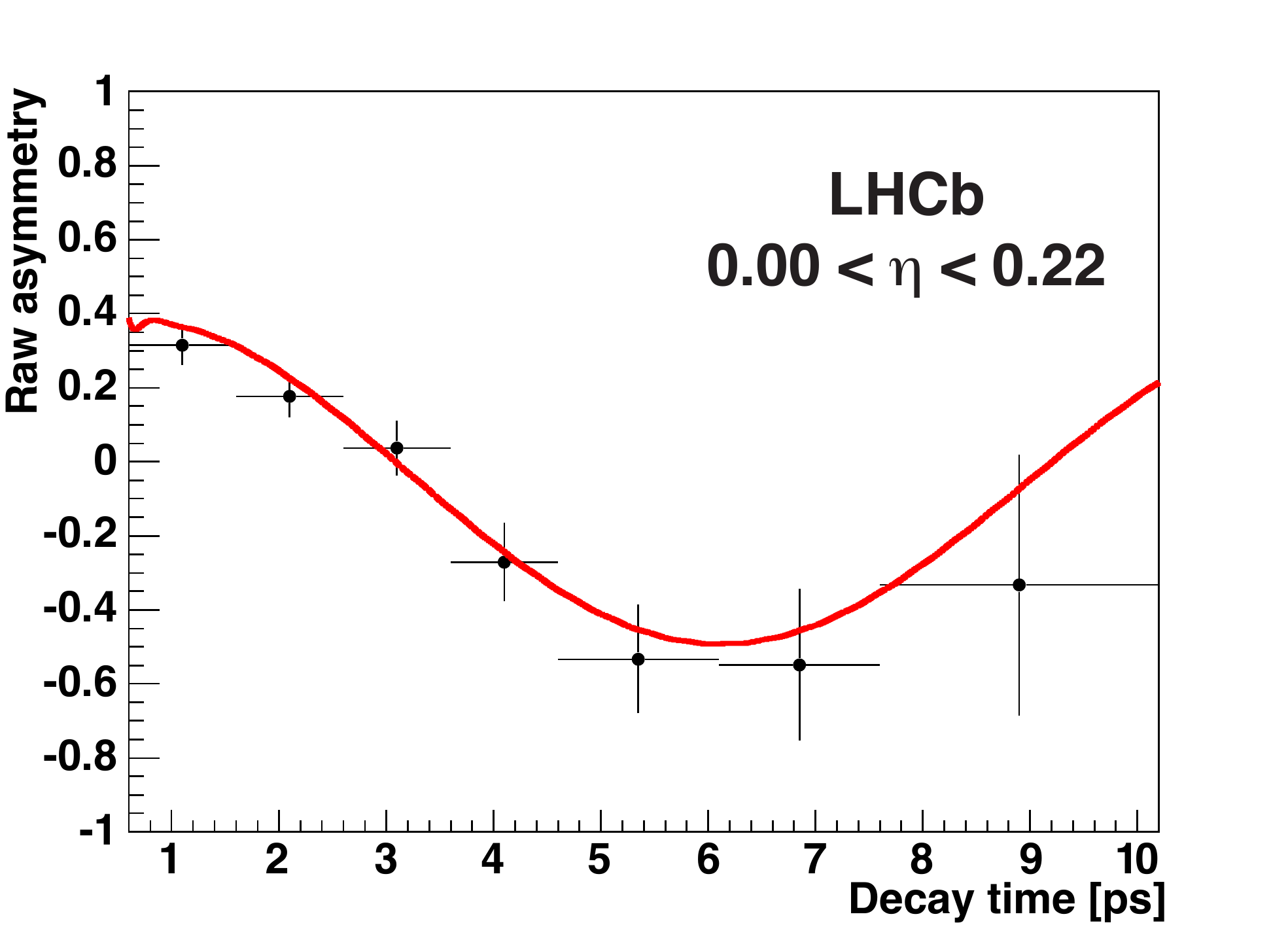}
    \includegraphics[width=0.45\textwidth]{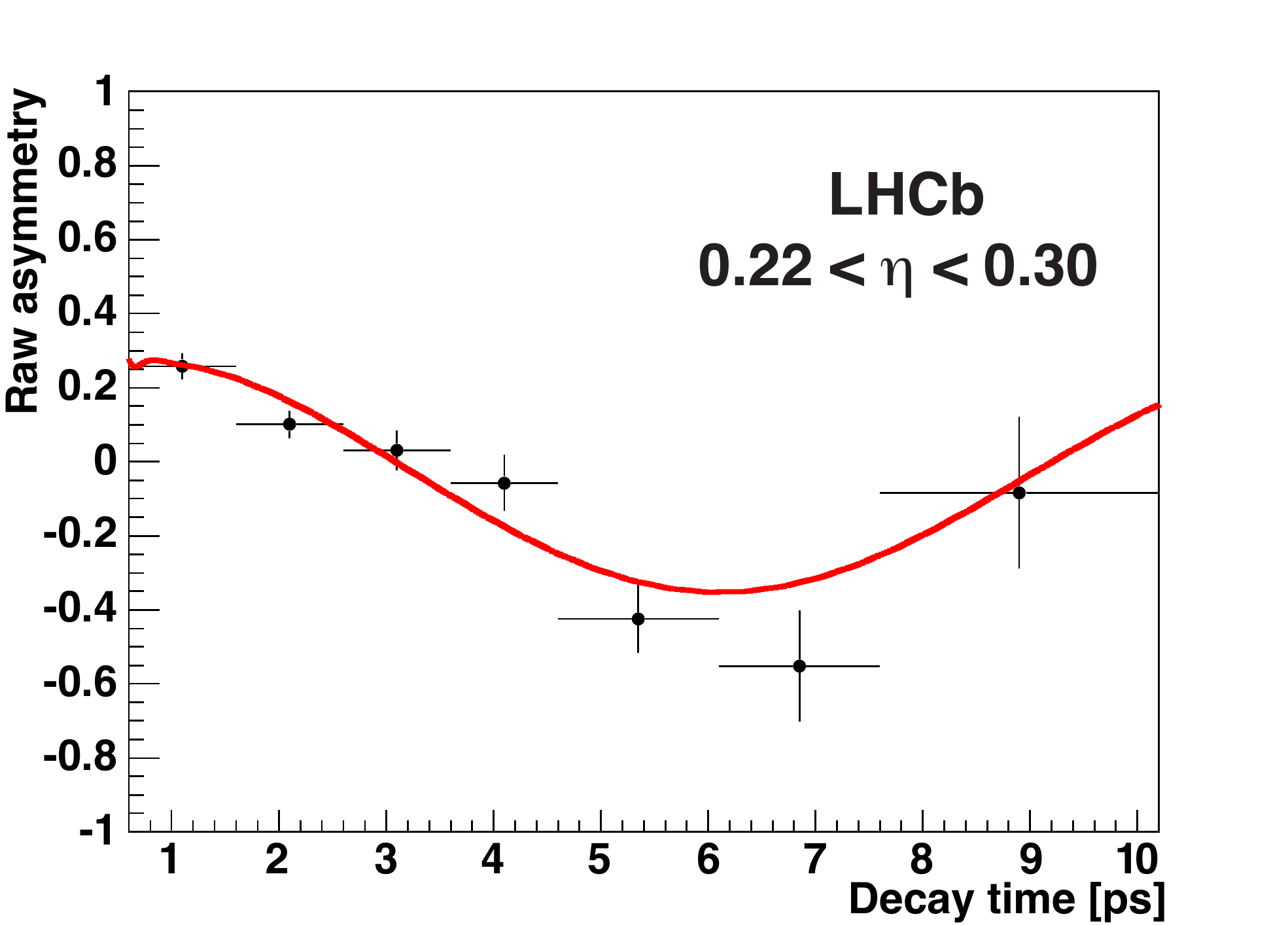}
    \includegraphics[width=0.45\textwidth]{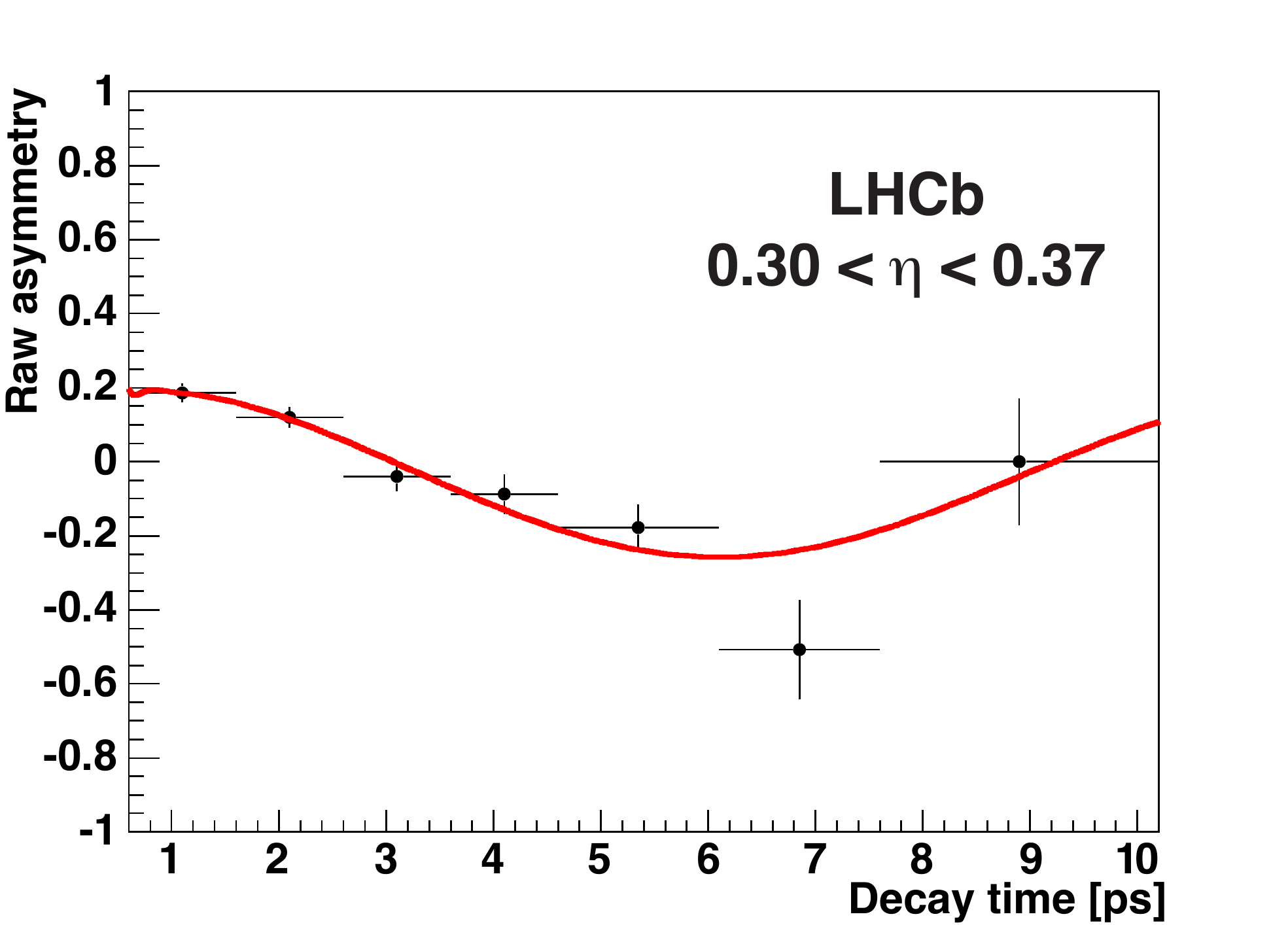}
    \includegraphics[width=0.45\textwidth]{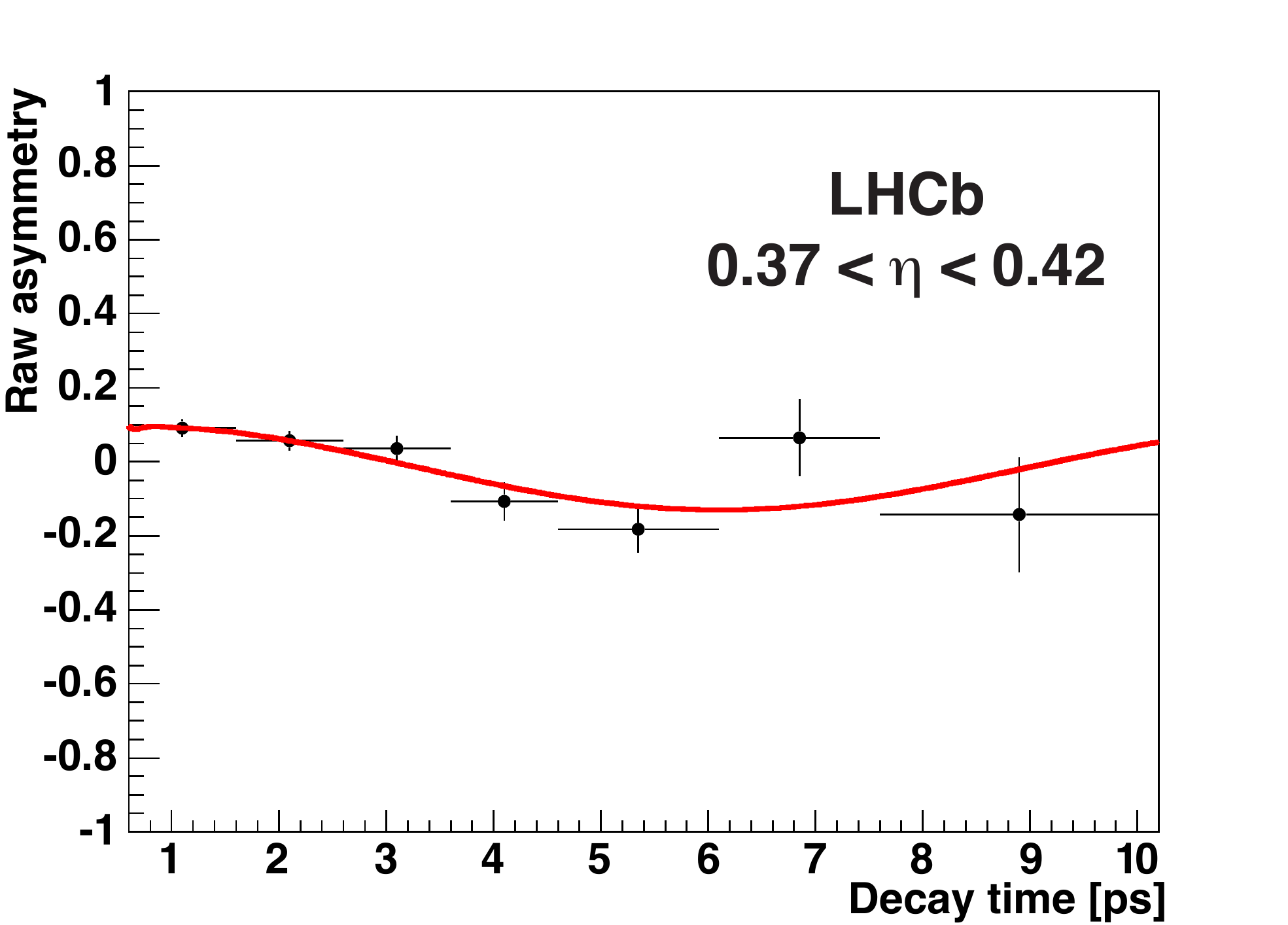}
    \includegraphics[width=0.45\textwidth]{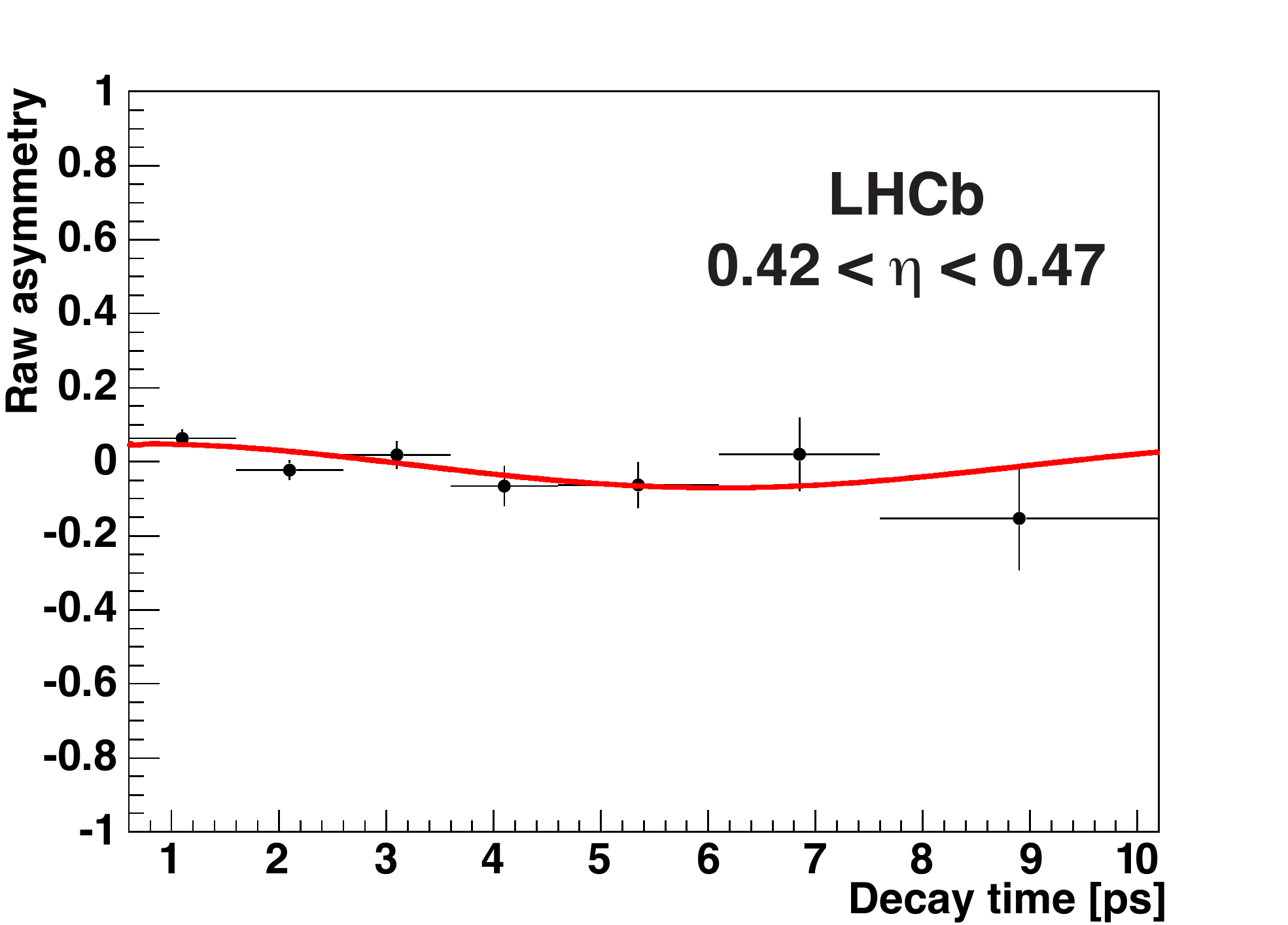}
 \end{center}
  \caption{\small Raw mixing asymmetries for candidates in the \BdToKpi signal mass region, corresponding to the five tagging categories, with the result of the fit overlaid.}
  \label{fig:rawMixingAsymmetry}
\end{figure}

To determine the values of $A_{k}^{\varepsilon}$, $\omega_{k}^\mathrm{tot}$ and $A_{k}^{\omega}$, we fit the model described in Sec.~\ref{sec:fitModel} to the $K^\pm\pi^\mp$ spectra.
In the $K^\pm\pi^\mp$ fit, the amount of \BdTopipi and \BsToKK cross-feed backgrounds below the \BdToKpi peak are fixed to the values obtained by performing a time-integrated simultaneous fit to all two-body invariant mass spectra, as in Ref.~\cite{LHCb-PAPER-2013-018}.
In Fig.~\ref{fig:bdkpiFit} the $K^\pm\pi^\mp$ invariant mass and decay time distributions are shown.
In Fig.~\ref{fig:rawMixingAsymmetry} the raw mixing asymmetry is shown for each of the five tagging categories, by considering only candidates with invariant mass in the region dominated by \BdToKpi decays, $5.20 < m <5.32$\gevcc. The asymmetry projection from the full fit is superimposed.
The $B^0 \to K^+\pi^-$ and $B^0_s \to K^-\pi^+$ event yields determined from the fit are $N(B^0 \to K^+\pi^-) = 49\hspace{0.5mm}356 \pm 335\,\mathrm{(stat)}$ and $N(B^0_s \to K^-\pi^+) = 3917 \pm 142\,\mathrm{(stat)}$, respectively. The mass differences are determined to be $\Delta m_{d} = 0.512 \pm 0.014\,\mathrm{(stat)}$\invps and $\Delta m_{s} = 17.84 \pm 0.11\,\mathrm{(stat)}$\invps. The $B^0$ and $B^0_s$ average lifetimes determined from the fit are $\tau(B^0)=1.523 \pm 0.007\,\mathrm{(stat)}$\ps and $\tau(B^0_s)=1.51 \pm 0.03\,\mathrm{(stat)}$\ps.
The signal tagging efficiencies and mistag probabilities are summarized in Table~\ref{tab:taggingPerformancesPIPI}. With the present precision, there is no evidence of non-zero asymmetries in the tagging efficiencies and mistag probabilities between $B^0_{(s)}$ and $\Bb^0_{(s)}$ mesons. The average effective tagging power is $\varepsilon_\mathrm{eff}=(2.45 \pm 0.25)\%$.
\begin{table}[t]
  \caption{\small Signal tagging efficiencies, mistag probabilities and associated asymmetries, corresponding to the five tagging categories, as determined from the $K^\pm\pi^\mp$ mass and decay time fit. The uncertainties are statististical only.}
  \begin{center}
\resizebox{1\textwidth}{!}{
    \begin{tabular}{c|c|c|c}
      Efficiency (\%) & Efficiency asymmetry (\%) &  Mistag probability (\%) & Mistag asymmetry (\%)\\
      \hline
      $\varepsilon_{1}^\mathrm{tot} = 1.92 \pm 0.06$ & $A_{1}^{\varepsilon} = -8 \pm 5$ &      $\omega_{1}^\mathrm{tot} = 20.0 \pm 2.8$ & $A_{1}^{\omega} = \phantom{-}0 \pm 10$ \\
      $\varepsilon_{2}^\mathrm{tot} = 4.07 \pm 0.09$ & $A_{2}^{\varepsilon} = \phantom{-}0 \pm 4$ &      $\omega_{2}^\mathrm{tot} = 28.3 \pm 2.0$ & $A_{2}^{\omega} = \phantom{-}5 \pm 5\phantom{0}$ \\
      $\varepsilon_{3}^\mathrm{tot} = 7.43 \pm 0.12$ & $A_{3}^{\varepsilon} = \phantom{-}2 \pm 3$ &      $\omega_{3}^\mathrm{tot} = 34.3 \pm 1.5$ & $A_{3}^{\omega} = -1 \pm 3\phantom{0}$ \\
      $\varepsilon_{4}^\mathrm{tot} = 7.90 \pm 0.13$ & $A_{4}^{\varepsilon} = -2 \pm 3$  &      $\omega_{4}^\mathrm{tot} = 41.9 \pm 1.5$ & $A_{4}^{\omega} = -2 \pm 2\phantom{0}$ \\
      $\varepsilon_{5}^\mathrm{tot} = 7.86 \pm 0.13$ & $A_{5}^{\varepsilon} = \phantom{-}0 \pm 3$ &      $\omega_{5}^\mathrm{tot} = 45.8 \pm 1.5$ & $A_{5}^{\omega} = -4 \pm 2\phantom{0}$
   \end{tabular}
}
  \end{center}
  \label{tab:taggingPerformancesPIPI}
\end{table}
From the fit, the production asymmetries for the \Bd and \Bs mesons are determined to be $A_\mathrm{P}\left(\Bd\right) = (0.6 \pm 0.9) \%$ and $A_\mathrm{P}\left(\Bs\right) = (7 \pm 5) \%$, where the uncertainties are statistical only.

\section{Results}
\label{sec:fitresults}

The fit to the mass and decay time distributions of the \BsToKK candidates determines the \CP asymmetry coefficients $C_{KK}$ and $S_{KK}$, whereas the \BdTopipi fit determines $C_{\pi\pi}$ and $S_{\pi\pi}$. In both fits, the yield of \BdToKpi cross-feed decays is fixed to the value obtained from a time-integrated fit, identical to that of Ref.~\cite{LHCb-PAPER-2013-018}. Furthermore, the flavour tagging efficiency asymmetries, mistag fractions and mistag asymmetries, and the \Bd and \Bs production asymmetries are constrained to the values measured in the fit described in the previous section, by multiplying the likelihood function with Gaussian terms.

\begin{figure}[t]
  \begin{center}
    \includegraphics[width=0.49\textwidth]{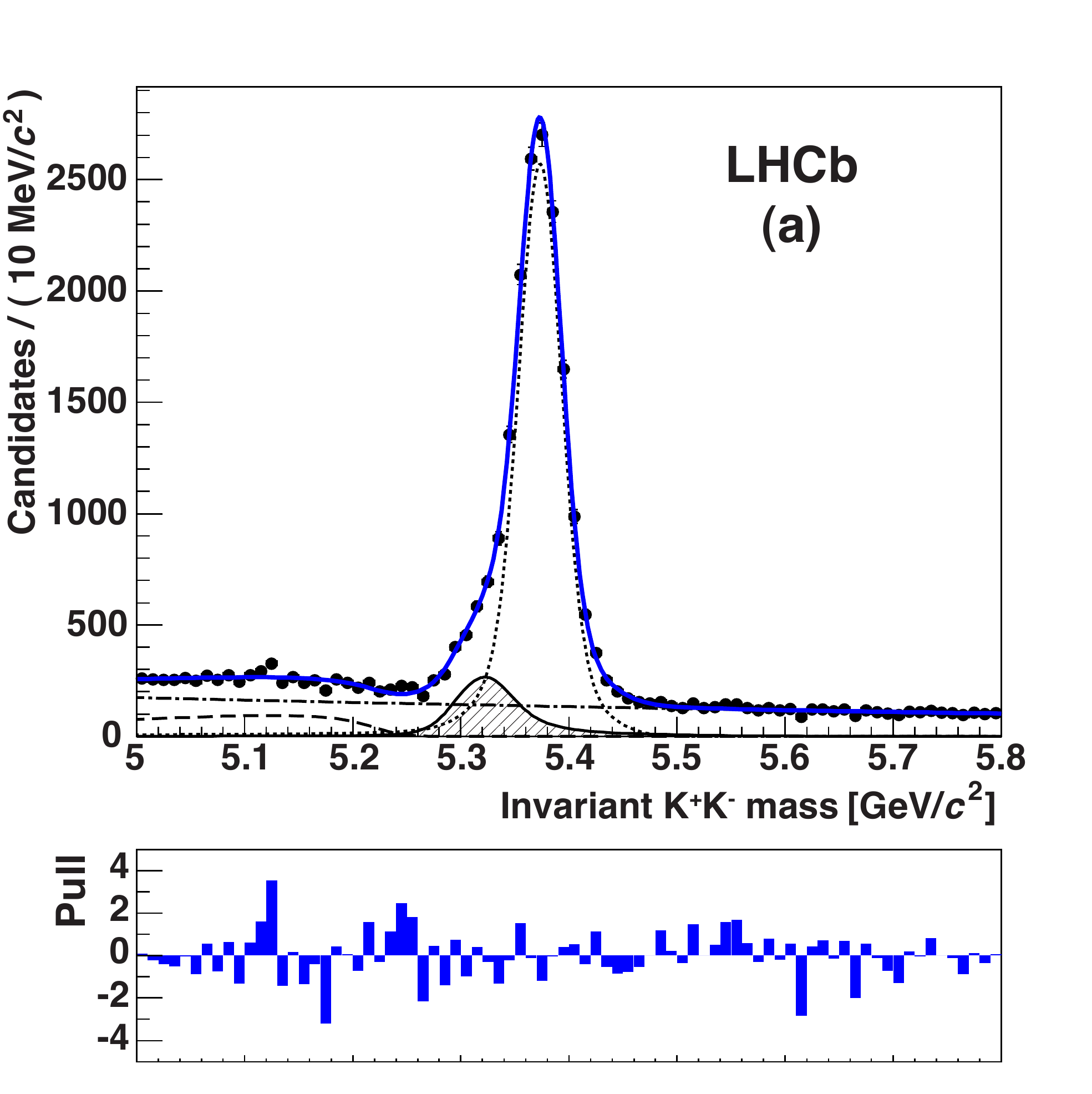}
    \includegraphics[width=0.49\textwidth]{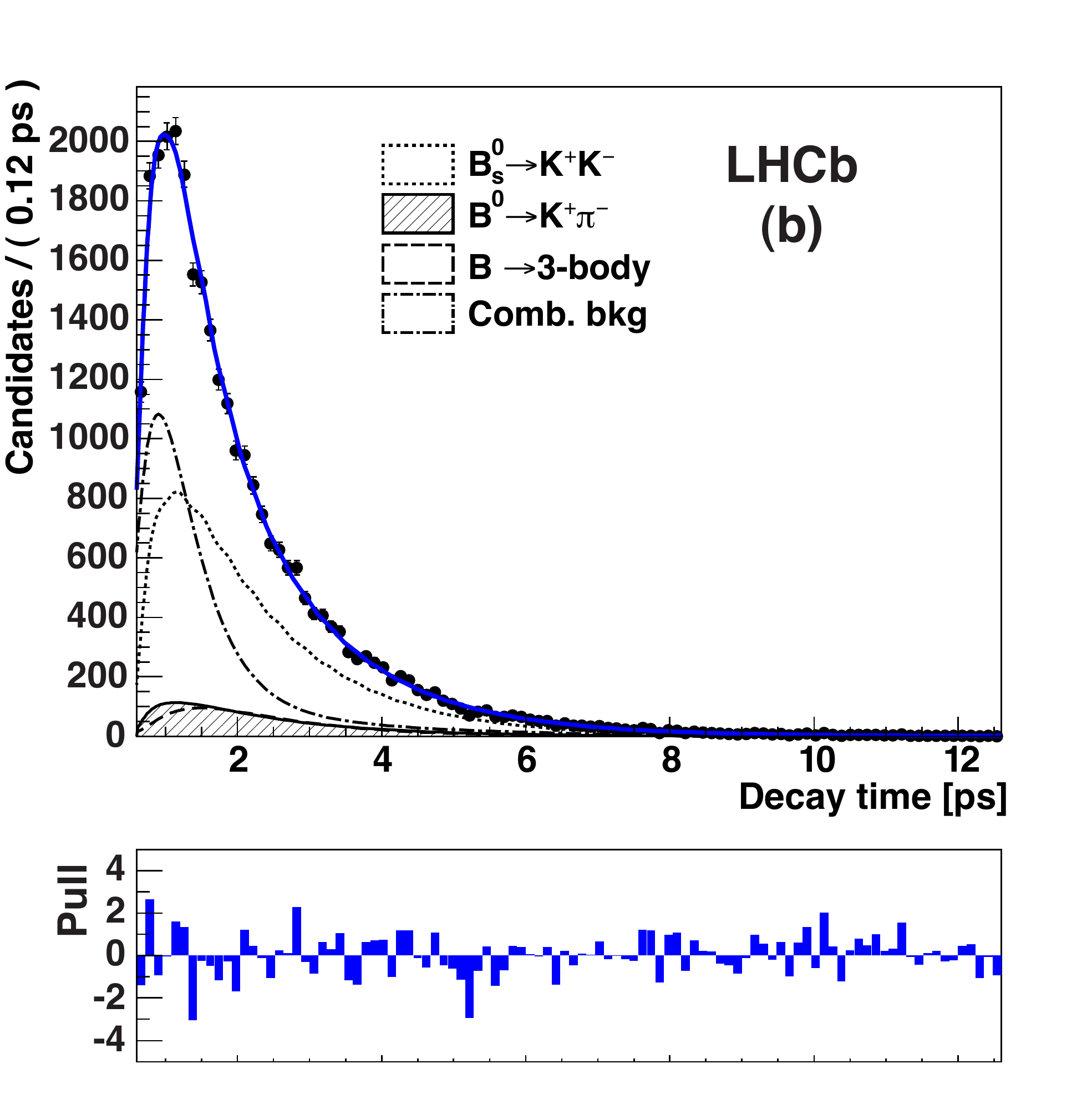}
  \end{center}
  \caption{\small Distributions of $K^+K^-$ (a) mass and (b) decay time, with the result of the fit overlaid. The main components contributing to the fit model are also shown.}
  \label{fig:bskkFit}
\end{figure}

\begin{figure}[t]
  \begin{center}
\includegraphics[width=0.7\textwidth]{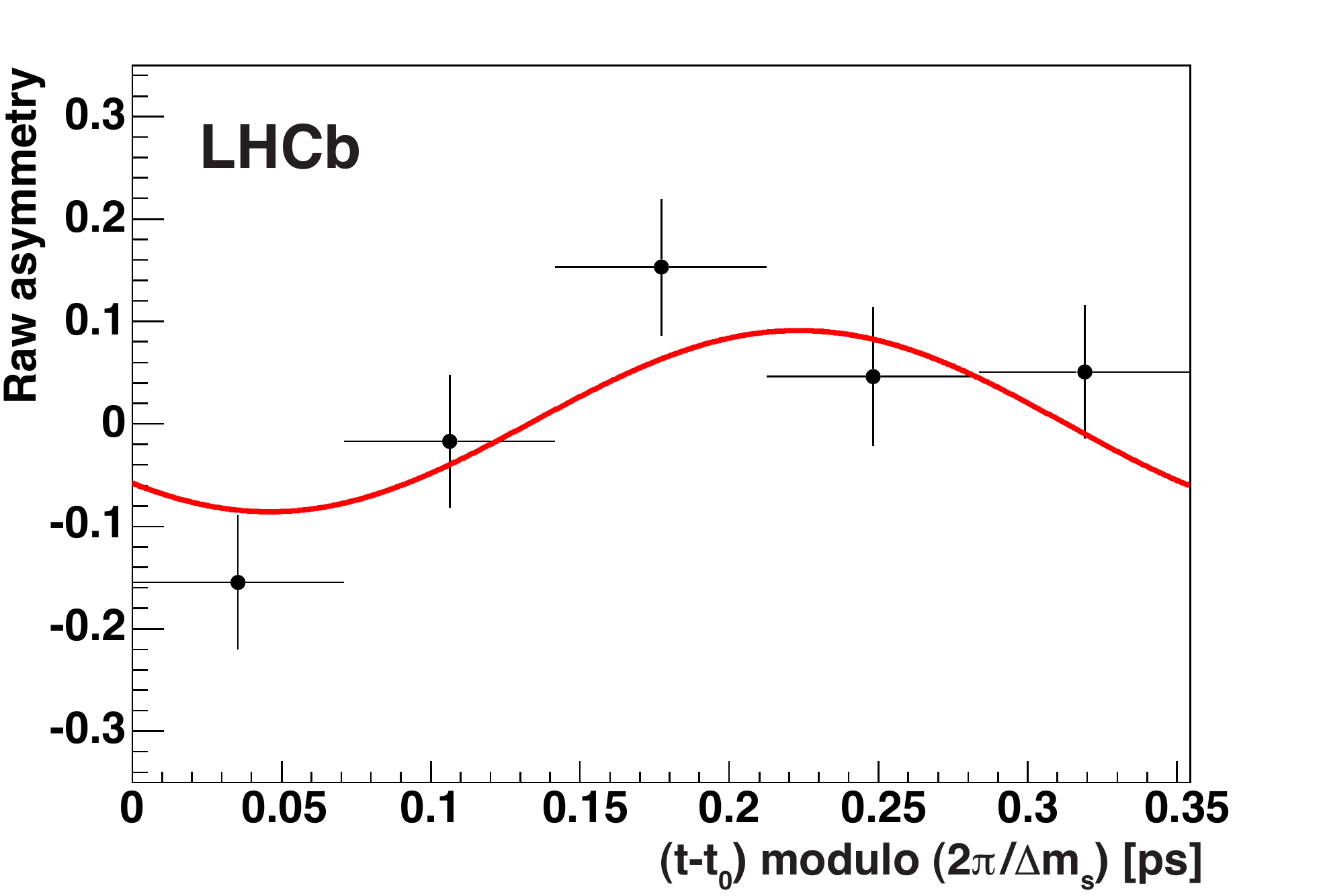}
  \end{center}
  \caption{\small Time-dependent raw asymmetry for candidates in the \BsToKK signal mass region with the result of the fit overlaid. In order to enhance the visibility of the oscillation, only candidates belonging to the first two tagging categories are used. The offset $t_0 = 0.6$\ps corresponds to the preselection requirement on the decay time.}
  \label{fig:rawCPAsymmetryKK}
\end{figure}

The $K^+K^-$ invariant mass and decay time distributions are shown in Fig.~\ref{fig:bskkFit}.
The raw time-dependent asymmetry is shown in Fig.~\ref{fig:rawCPAsymmetryKK} for candidates with invariant mass in the region dominated by signal events, $5.30 < m < 5.44$\gevcc, and belonging to the first two tagging categories.
The $B^0_s \to K^+K^-$ event yield is determined to be $N(B^0_s \to K^+K^-) = 14\hspace{0.5mm}646 \pm 159\,(\mathrm{stat})$, while the $B^0_s$ decay width difference from the fit is $\Delta\Gamma_s = 0.104 \pm 0.016\,\mathrm{(stat)}$\invps.
The values of $C_{KK}$ and $S_{KK}$ are determined to be
\begin{equation}
  C_{KK} = 0.14 \pm 0.11\,\mathrm{(stat)},\qquad S_{KK} = 0.30 \pm 0.12\,\mathrm{(stat)},\nonumber
\end{equation}
with correlation coefficient $\rho\left(C_{KK},\,S_{KK}\right) = 0.02$. The small value of the correlation coefficient is a consequence of the large $B^0_s$ mixing frequency. 
An alternative fit, fixing the value of $\Delta\Gamma_{s}$  to $0.106$\invps~\cite{LHCb-PAPER-2013-002} and leaving $A^{\Delta\Gamma}_{KK}$ free to vary, is also performed as a cross-check. Central values and statistical uncertainties of $C_{KK}$ and $S_{KK}$ are almost unchanged, and $A^{\Delta\Gamma}_{KK}$ is determined to be $0.91 \pm 0.08\,\mathrm{(stat)}$.

\begin{figure}[t]
  \begin{center}
    \includegraphics[width=0.49\textwidth]{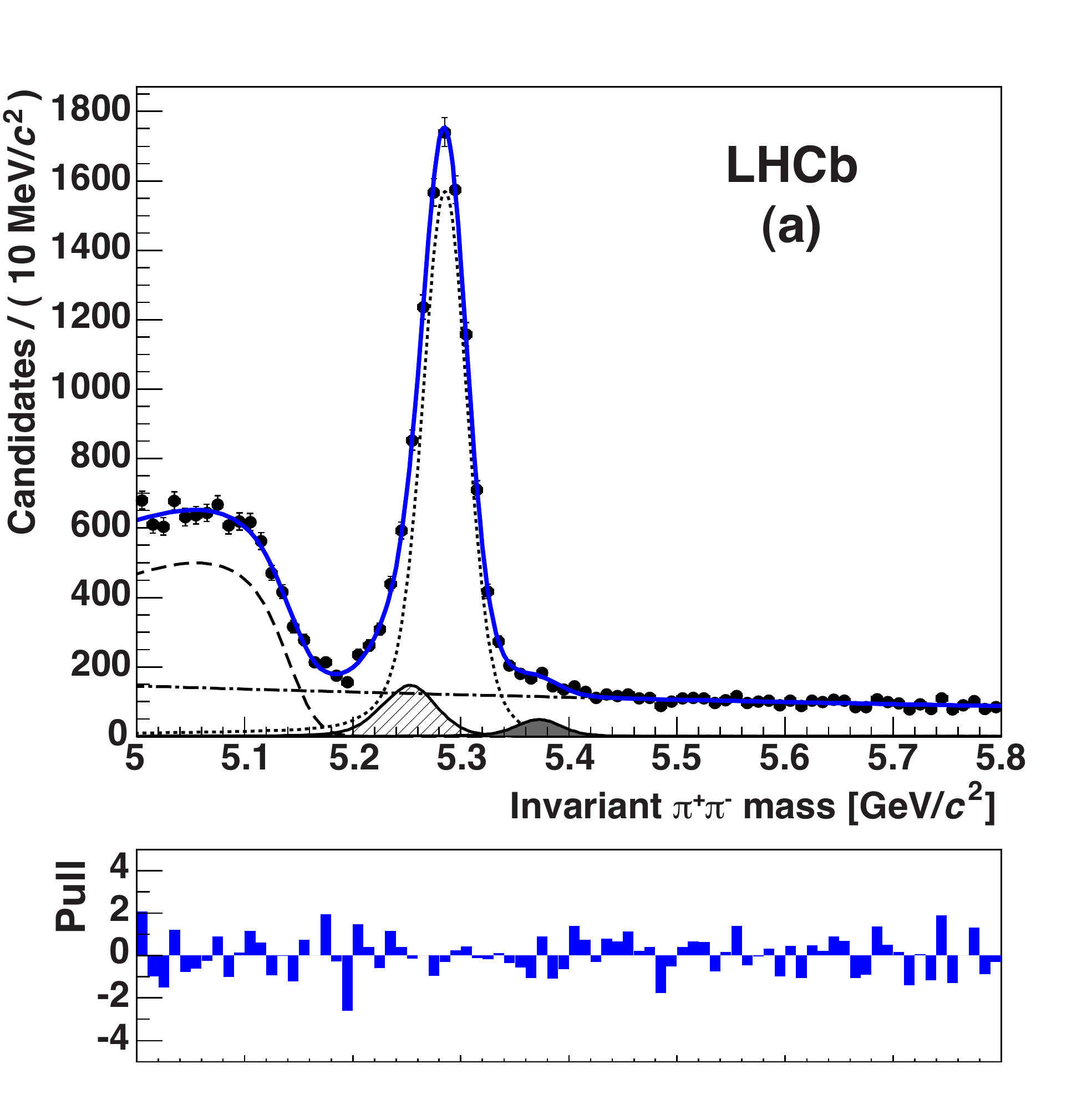}
   \includegraphics[width=0.49\textwidth]{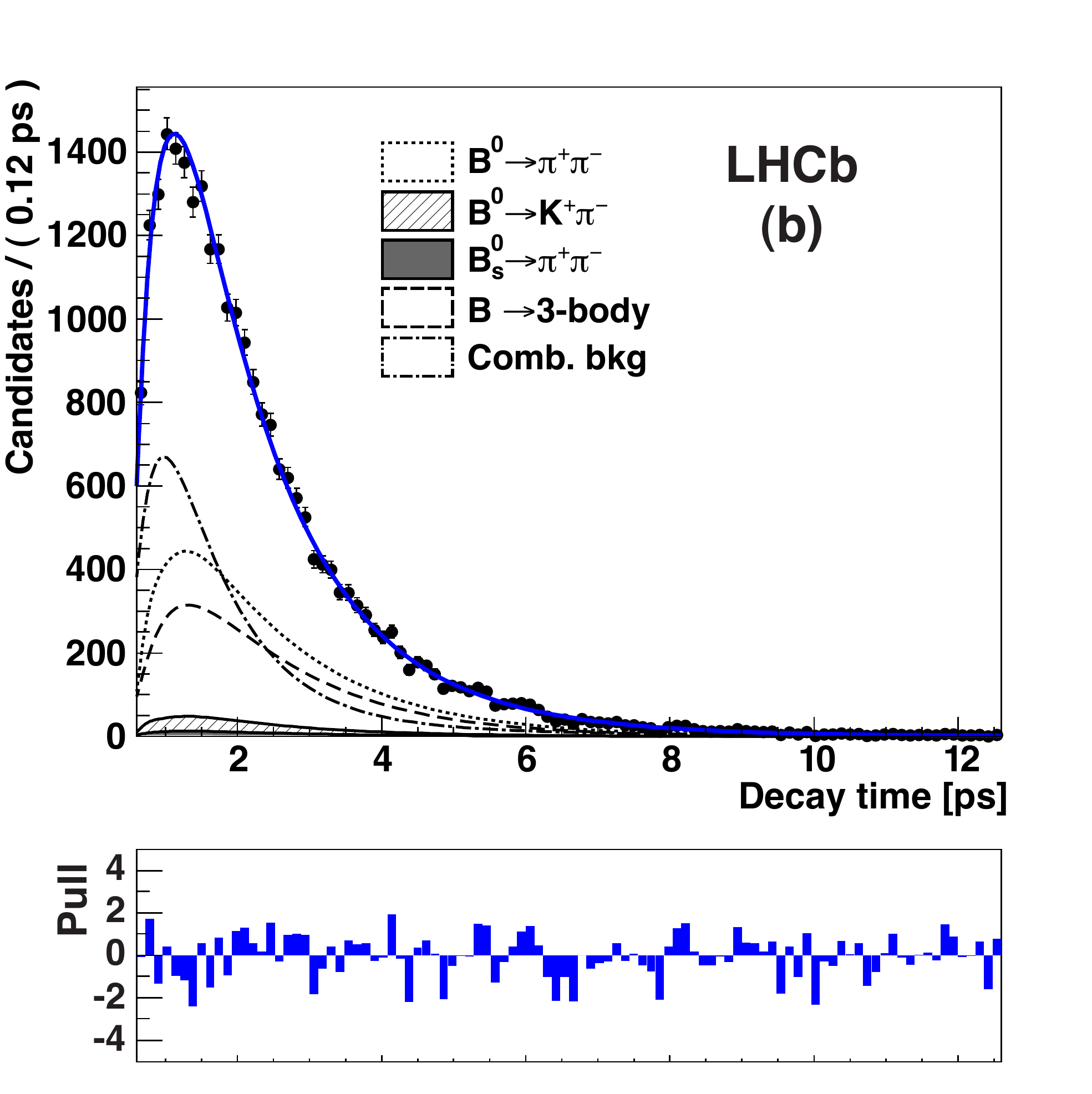}
  \end{center}
  \caption{\small Distributions of $\pi^+\pi^-$ (a) mass and (b) decay time, with the result of the fit overlaid. The main components contributing to the fit model are also shown.}
  \label{fig:bdpipiFit}
\end{figure}

\begin{figure}[t]
  \begin{center}
\includegraphics[width=0.7\textwidth]{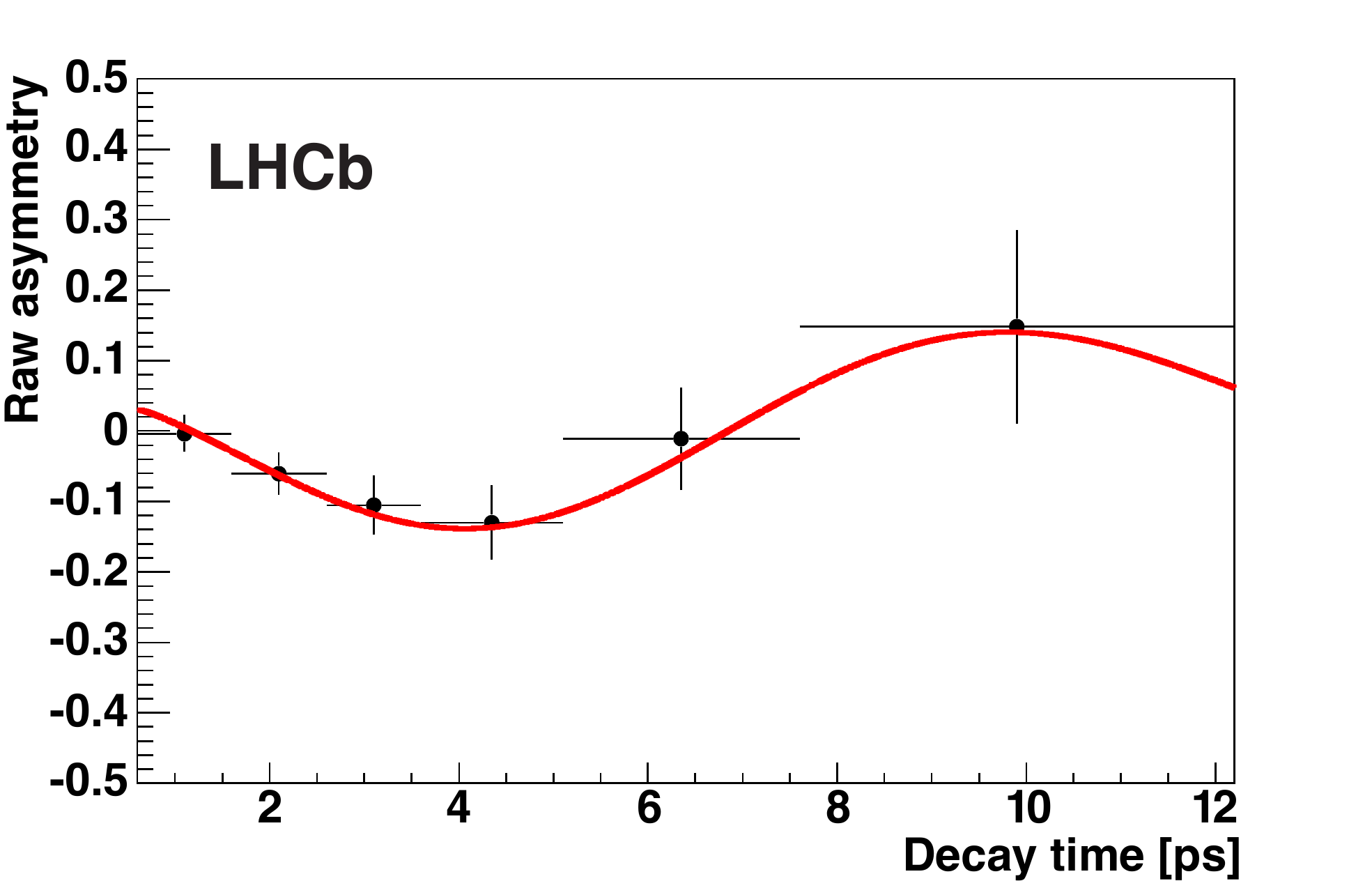}
  \end{center}
  \caption{\small Time-dependent raw asymmetry for candidates in the \BdTopipi signal mass region with the result of the fit overlaid. Tagged candidates belonging to all tagging categories are used.}
  \label{fig:rawCPAsymmetryPIPI}
\end{figure}

Although very small, a component accounting for the presence of the $B_{s}^{0}\to \pi^{+}\pi^{-}$ decay~\cite{LHCb-PAPER-2012-002} is introduced in the \BdTopipi fit. This component is described using the signal model, but assuming no \CP violation.
The $\pi^+\pi^-$ invariant mass and decay time distributions are shown in Fig.~\ref{fig:bdpipiFit}.
The raw time-dependent asymmetry is shown in Fig.~\ref{fig:rawCPAsymmetryPIPI} for candidates with invariant mass in the region dominated by signal events, $5.20 < m < 5.36$\gevcc.
The $B^0 \to \pi^+\pi^-$ event yield is determined to be $N(B^0 \to \pi^+\pi^-) = 9170 \pm 144\,\mathrm{(stat)}$, while the $B^0$ average lifetime from the fit is $\tau(B^0) = 1.55 \pm 0.02\,\mathrm{(stat)}$\ps.
The values of $C_{\pi\pi}$ and $S_{\pi\pi}$ are determined to be
\begin{equation}
  C_{\pi\pi} = -0.38 \pm 0.15\,\mathrm{(stat)},\qquad S_{\pi\pi} = -0.71 \pm 0.13\,\mathrm{(stat)},\nonumber
\end{equation}
with correlation coefficient $\rho\left(C_{\pi\pi},\,S_{\pi\pi}\right) = 0.38$.

\clearpage

\section{Systematic uncertainties}\label{sec:systematics}

Several sources of systematic uncertainty that may affect the determination of the direct and mixing-induced \CP-violating asymmetries in $B^0_s \to K^+K^-$ and $B^0 \to \pi^+\pi^-$ decays are considered. For the invariant mass model, the accuracy of PID efficiencies and the description of mass shapes for all components (signals, combinatorial, partially reconstructed three-body and cross-feed backgrounds) are investigated. For the decay time model, systematic effects related to the decay time resolution and acceptance are studied. The effects of the external input variables used in the fits ($\Delta m_{s}$, $\Delta m_{d}$, $\Delta\Gamma_{s}$ and $\Gamma_{s}$), and the parameterization of the backgrounds are also considered.
To estimate the contribution of each single source the fit is repeated after having modified the baseline parameterization. The shifts from the relevant baseline values are accounted for as systematic uncertainties.

The PID efficiencies are used to compute the yields of cross-feed backgrounds present in the $K^\pm \pi^\mp$, $\pi^+\pi^-$ and $K^+K^-$ mass distributions. In order to estimate the impact of imperfect PID calibration, unbinned maximum likelihood fits are performed after having altered the number of cross-feed background events present in the relevant mass spectra, according to the systematic uncertainties associated to the PID efficiencies.

An estimate of the uncertainty due to possible mismodelling of the final-state radiation is determined by varying the amount of emitted
radiation~\cite{Baracchini:2005wp} in the signal shape parameterization, according to studies performed on simulated events, in which final state radiation is generated using \photos~\cite{Golonka:2005pn}.
The possibility of an incorrect description of the signal mass model is investigated by replacing the double Gaussian function with the sum of three Gaussian functions, where the third component has fixed fraction ($5\%$) and width ($50$\mevcc), and is aimed at describing long tails, as observed in simulation.
The systematic uncertainties related to the parameterization of the invariant mass shape for the combinatorial background are investigated by replacing the exponential shape with a straight line function. 
For the case of the cross-feed backgrounds, two distinct systematic uncertainties are estimated: one due to a relative bias in the mass scale of the simulated distributions with respect to the signal distributions in data, and another to account for the difference in mass resolution between simulation and data.

Systematic uncertainties associated to the decay time resolution are investigated by altering the resolution model in different ways. The width of the single Gaussian model used in the baseline fit is changed by $\pm 10$\fs. Effects due to a possible bias in the decay time measurement are accounted for by repeating the fit with a bias of $\pm 2$\fs. Finally, the single Gaussian model is substituted by a triple Gaussian model, where the fractions of the Gaussian functions are taken from simulation and the widths are rescaled to match the average width of $50$\fs used in the baseline fit.

To estimate systematic uncertainties arising from the choice of parameterization for backgrounds, fits with alternative parameterizations are performed.
To account for possible inaccuracies in the decay time acceptances determined from simulation, the fits are repeated fixing $\Gamma_{d}$ to $0.658$\invps and $\Delta\Gamma_s$ to $0.106$\invps, and leaving the acceptance parameters $p_i$ free to vary.

Systematic uncertainties related to the use of external inputs are estimated by varying the input quantities by $\pm 1\sigma$ of the corresponding measurements. In particular, this is done in the $B^0 \to K^+\pi^-$ and $B^0_s \to K^-\pi^+$ fit for $\Delta\Gamma_{s}$~($\pm 0.013$\invps), in the $B^0 \to \pi^+\pi^-$ fit for $\Delta m_{d}$~($\pm 0.006$\invps), and in the $B^0_s \to K^+K^-$ fit for $\Delta m_{s}$~($\pm 0.024$\invps) and $\Gamma_{s}$~($\pm 0.007$\invps).

Following the procedure outlined above, we also estimate the systematic uncertainties affecting the flavour tagging efficiencies, mistag probabilities and production asymmetries, and propagate these uncertainties to the systematic uncertainties on the direct and mixing-induced \CP asymmetry coefficients in $B^0_s \to K^+K^-$ and $B^0 \to \pi^+\pi^-$ decays.
The final systematic uncertainties on these coefficients are summarized in Table~\ref{tab:systematicsKK}. They turn out to be much smaller than the corresponding statistical uncertainties reported in Sec.~\ref{sec:fitresults}.

\begin{table}[t]
  \caption{\small Systematic uncertainties affecting the $B^0_s \to K^+K^-$ and $B^0 \to \pi^+\pi^-$ direct and mixing-induced \CP asymmetry coefficients. The total systematic uncertainties are obtained by summing the individual contributions in quadrature.}
  \begin{center}
\resizebox{1\textwidth}{!}{
    \begin{tabular}{rl|cc|cc}
      \multicolumn{2}{c|}{Systematic uncertainty}  & $C_{KK}$ & $S_{KK}$  & $C_{\pi\pi}$ & $S_{\pi\pi}$ \\
      \hline
      \multicolumn{2}{c|}{Particle identification}                                                       & $0.003$ & $0.003$                                                        & $0.002$ & $0.004$ \\

     \multicolumn{2}{c|}{Flavour tagging}                                                             & $0.008$ & $0.009$                            & $0.010$ & $0.011$ \\
      \multicolumn{2}{c|}{Production asymmetry}                                                   & $0.002$ & $0.002$                                                & $0.003$ & $0.002$ \\
      \hline
     \multirow{2}{*}{Signal mass:} & final state radiation                                        & $0.002$ & $0.001$                                        & $0.001$ & $0.002$ \\
                         & shape model                                             & $0.003$ & $0.004$                                            & $0.001$ & $0.004$ \\
\hline
      \multirow{2}{*}{Bkg. mass:} & combinatorial               & $<0.001\phantom{11}$& $<0.001\phantom{11}$              & $<0.001\phantom{11}$ & $<0.001\phantom{11}$ \\
       & cross-feed                                     & $0.002$ & $0.003$                                     & $0.002$ & $0.004$ \\
\hline
     \multirow{4}{*}{Sig. decay time:}& acceptance                                                      & $0.010$ & $0.018$                                                      & $0.002$ & $0.003$ \\
     & resolution width                                             & $0.020$ & $0.025$                                             & $<0.001\phantom{11}$ & $<0.001\phantom{11}$ \\
     & resolution bias                                               & $0.009$ & $0.007$                                               & $<0.001\phantom{11}$ & $<0.001\phantom{11}$ \\
     & resolution model                                            & $0.008$ & $0.015$                                           & $<0.001\phantom{11}$ & $<0.001\phantom{11}$ \\
\hline
      \multirow{3}{*}{Bkg. decay time:} & cross-feed                                            & $<0.001\phantom{11}$ & $<0.001\phantom{11}$                                           & $0.005$ & $0.002$ \\
      & combinatorial                   & $0.008$ & $0.006$                  & $0.015$ & $0.011$ \\
     & three-body                          & $0.001$ & $0.003$                         & $0.003$ & $0.005$ \\
\hline
      \multirow{3}{*}{Ext. inputs:} & $\Delta m_s$                                                & $0.015$ & $0.018$ & - & - \\
       & $\Delta m_d$                                               & - & - & $0.013$ & $0.010$ \\
      & $\Gamma_{s}$                                            & $0.004$ & $0.005$ & - & - \\
     \hline
      \multicolumn{2}{c|}{Total}                                                                                 & $0.032$ & $0.042$                                                                        & $0.023$ & $0.021$
    \end{tabular}
}
  \end{center}
  \label{tab:systematicsKK}
\end{table}

\section{Conclusions}
\label{sec:conclusions}

The measurement of time-dependent \CP violation in $B^0_s \to K^+K^-$ and $B^0 \to \pi^+\pi^-$ decays, based on a data
sample corresponding to an integrated luminosity of 1.0~fb$^{-1}$, has been presented. 
The results for the $B^0_s \to K^+K^-$ decay are 
\begin{equation}
\begin{split}
C_{KK}=0.14 \pm 0.11\,\mathrm{(stat)} \pm 0.03\,\mathrm{(syst)},\nonumber\\
S_{KK}=0.30 \pm 0.12\,\mathrm{(stat)} \pm 0.04\,\mathrm{(syst)},
\end{split}
\end{equation}
with a statistical correlation coefficient of $0.02$.
The results for the  $B^0 \to \pi^+\pi^-$ decay are
\begin{equation}
\begin{split}
C_{\pi\pi}=-0.38 \pm 0.15\,\mathrm{(stat)} \pm 0.02\,\mathrm{(syst)},\nonumber\\
S_{\pi\pi}=-0.71 \pm 0.13\,\mathrm{(stat)} \pm 0.02\,\mathrm{(syst)},
\end{split}
\end{equation}
with a statistical correlation coefficient of $0.38$.

Dividing the central values of the measurements by the sum in quadrature of statistical and systematic uncertainties, and taking correlations into account, the significances for $(C_{KK},\,S_{KK})$ and $(C_{\pi\pi},\,S_{\pi\pi})$ to differ from $(0,\,0)$ are determined to be 2.7$\sigma$ and 5.6$\sigma$, respectively. The parameters $C_{KK}$ and $S_{KK}$ are measured for the first time. The measurements of $C_{\pi\pi}$ and $S_{\pi\pi}$ are in good agreement with previous measurements by BaBar~\cite{Lees:2013bb} and Belle~\cite{Adachi:2013mae}, and those of $C_{KK}$ and  $S_{KK}$ are compatible with theoretical SM predictions~\cite{Beneke:2003zv,DescotesGenon:2006wc,Chiang:2006ih,Fleischer:2010ib}.
These results, together with those from BaBar and Belle, allow the determination of the unitarity triangle angle $\gamma$ using decays affected by penguin processes~\cite{Fleischer:1999pa,Ciuchini:2012gd}. The comparison to the value of $\gamma$ determined from tree-level decays will provide a test of the SM and constrain possible non-SM contributions.

\clearpage

%% file: acknowledgements.tex
\section*{Acknowledgements}

\noindent We express our gratitude to our colleagues in the CERN
accelerator departments for the excellent performance of the LHC. We
thank the technical and administrative staff at the LHCb
institutes. We acknowledge support from CERN and from the national
agencies: CAPES, CNPq, FAPERJ and FINEP (Brazil); NSFC (China);
CNRS/IN2P3 and Region Auvergne (France); BMBF, DFG, HGF and MPG
(Germany); SFI (Ireland); INFN (Italy); FOM and NWO (The Netherlands);
SCSR (Poland); MEN/IFA (Romania); MinES, Rosatom, RFBR and NRC
``Kurchatov Institute'' (Russia); MinECo, XuntaGal and GENCAT (Spain);
SNSF and SER (Switzerland); NAS Ukraine (Ukraine); STFC (United
Kingdom); NSF (USA). We also acknowledge the support received from the
ERC under FP7. The Tier1 computing centres are supported by IN2P3
(France), KIT and BMBF (Germany), INFN (Italy), NWO and SURF (The
Netherlands), PIC (Spain), GridPP (United Kingdom). We are thankful
for the computing resources put at our disposal by Yandex LLC
(Russia), as well as to the communities behind the multiple open
source software packages that we depend on.

%% file: main.bbl
\ifx\mcitethebibliography\mciteundefinedmacro
\PackageError{LHCb.bst}{mciteplus.sty has not been loaded}
{This bibstyle requires the use of the mciteplus package.}\fi
\providecommand{\href}[2]{#2}
\begin{mcitethebibliography}{10}
\mciteSetBstSublistMode{n}
\mciteSetBstMaxWidthForm{subitem}{\alph{mcitesubitemcount})}
\mciteSetBstSublistLabelBeginEnd{\mcitemaxwidthsubitemform\space}
{\relax}{\relax}

\bibitem{Cabibbo:1963yz}
N.~Cabibbo, \ifthenelse{\boolean{articletitles}}{{\it {Unitary symmetry and
  leptonic decays}},
  }{}\href{http://dx.doi.org/10.1103/PhysRevLett.10.531}{Phys.\ Rev.\ Lett.\
  {\bf 10} (1963) 531}\relax
\mciteBstWouldAddEndPuncttrue
\mciteSetBstMidEndSepPunct{\mcitedefaultmidpunct}
{\mcitedefaultendpunct}{\mcitedefaultseppunct}\relax
\EndOfBibitem
\bibitem{Kobayashi:1973fv}
M.~Kobayashi and T.~Maskawa, \ifthenelse{\boolean{articletitles}}{{\it {CP
  violation in the renormalizable theory of weak interaction}},
  }{}\href{http://dx.doi.org/10.1143/PTP.49.652}{Prog.\ Theor.\ Phys.\  {\bf
  49} (1973) 652}\relax
\mciteBstWouldAddEndPuncttrue
\mciteSetBstMidEndSepPunct{\mcitedefaultmidpunct}
{\mcitedefaultendpunct}{\mcitedefaultseppunct}\relax
\EndOfBibitem
\bibitem{Fleischer:1999pa}
R.~Fleischer, \ifthenelse{\boolean{articletitles}}{{\it {New strategies to
  extract $\beta$ and $\gamma$ from $B_d \to \pi^+ \pi^-$ and $B_s \to K^+
  K^-$}}, }{}\href{http://dx.doi.org/10.1016/S0370-2693(99)00640-1}{Phys.\
  Lett.\  {\bf B459} (1999) 306},
  \href{http://arxiv.org/abs/hep-ph/9903456}{{\tt arXiv:hep-ph/9903456}}\relax
\mciteBstWouldAddEndPuncttrue
\mciteSetBstMidEndSepPunct{\mcitedefaultmidpunct}
{\mcitedefaultendpunct}{\mcitedefaultseppunct}\relax
\EndOfBibitem
\bibitem{Gronau:2000md}
M.~Gronau and J.~L. Rosner, \ifthenelse{\boolean{articletitles}}{{\it {The role
  of $B_s \to K\pi$ in determining the weak phase $\gamma$}},
  }{}\href{http://dx.doi.org/10.1016/S0370-2693(00)00508-6}{Phys.\ Lett.\  {\bf
  B482} (2000) 71}, \href{http://arxiv.org/abs/hep-ph/0003119}{{\tt
  arXiv:hep-ph/0003119}}\relax
\mciteBstWouldAddEndPuncttrue
\mciteSetBstMidEndSepPunct{\mcitedefaultmidpunct}
{\mcitedefaultendpunct}{\mcitedefaultseppunct}\relax
\EndOfBibitem
\bibitem{Lipkin:2005pb}
H.~J. Lipkin, \ifthenelse{\boolean{articletitles}}{{\it {Is observed direct CP
  violation in $B_d \to K^+\pi^-$ due to new physics? Check standard model
  prediction of equal violation in $B_s \to K^-\pi^+$}},
  }{}\href{http://dx.doi.org/10.1016/j.physletb.2005.06.023}{Phys.\ Lett.\
  {\bf B621} (2005) 126}, \href{http://arxiv.org/abs/hep-ph/0503022}{{\tt
  arXiv:hep-ph/0503022}}\relax
\mciteBstWouldAddEndPuncttrue
\mciteSetBstMidEndSepPunct{\mcitedefaultmidpunct}
{\mcitedefaultendpunct}{\mcitedefaultseppunct}\relax
\EndOfBibitem
\bibitem{Fleischer:2007hj}
R.~Fleischer, \ifthenelse{\boolean{articletitles}}{{\it {$B_{s,d} \to \pi
  \pi,\, \pi K,\, KK$: status and prospects}},
  }{}\href{http://dx.doi.org/10.1140/epjc/s10052-007-0391-7}{Eur.\ Phys.\ J.\
  {\bf C52} (2007) 267}, \href{http://arxiv.org/abs/0705.1121}{{\tt
  arXiv:0705.1121}}\relax
\mciteBstWouldAddEndPuncttrue
\mciteSetBstMidEndSepPunct{\mcitedefaultmidpunct}
{\mcitedefaultendpunct}{\mcitedefaultseppunct}\relax
\EndOfBibitem
\bibitem{Fleischer:2010ib}
R.~Fleischer and R.~Knegjens, \ifthenelse{\boolean{articletitles}}{{\it {In
  pursuit of new physics with $B^0_s \to K^+K^-$}},
  }{}\href{http://dx.doi.org/10.1140/epjc/s10052-010-1532-y}{Eur.\ Phys.\ J.\
  {\bf C71} (2011) 1532}, \href{http://arxiv.org/abs/1011.1096}{{\tt
  arXiv:1011.1096}}\relax
\mciteBstWouldAddEndPuncttrue
\mciteSetBstMidEndSepPunct{\mcitedefaultmidpunct}
{\mcitedefaultendpunct}{\mcitedefaultseppunct}\relax
\EndOfBibitem
\bibitem{Gronau:1990ka}
M.~Gronau and D.~London, \ifthenelse{\boolean{articletitles}}{{\it {Isospin
  analysis of CP asymmetries in B decays}},
  }{}\href{http://dx.doi.org/10.1103/PhysRevLett.65.3381}{Phys.\ Rev.\ Lett.\
  {\bf 65} (1990) 3381}\relax
\mciteBstWouldAddEndPuncttrue
\mciteSetBstMidEndSepPunct{\mcitedefaultmidpunct}
{\mcitedefaultendpunct}{\mcitedefaultseppunct}\relax
\EndOfBibitem
\bibitem{Ciuchini:2012gd}
M.~Ciuchini, E.~Franco, S.~Mishima, and L.~Silvestrini,
  \ifthenelse{\boolean{articletitles}}{{\it {Testing the Standard Model and
  searching for new physics with $B_d \to \pi \pi$ and $B_s \to K K$ decays}},
  }{}\href{http://dx.doi.org/10.1007/JHEP10(2012)029}{JHEP {\bf 10} (2012)
  029}, \href{http://arxiv.org/abs/1205.4948}{{\tt arXiv:1205.4948}}\relax
\mciteBstWouldAddEndPuncttrue
\mciteSetBstMidEndSepPunct{\mcitedefaultmidpunct}
{\mcitedefaultendpunct}{\mcitedefaultseppunct}\relax
\EndOfBibitem
\bibitem{LHCb-PAPER-2011-029}
LHCb collaboration, R.~Aaij {\em et~al.},
  \ifthenelse{\boolean{articletitles}}{{\it {First evidence of direct \CP
  violation in charmless two-body decays of \Bs mesons}},
  }{}\href{http://dx.doi.org/10.1103/PhysRevLett.108.201601}{Phys.\ Rev.\
  Lett.\  {\bf 108} (2012) 201601}, \href{http://arxiv.org/abs/1202.6251}{{\tt
  arXiv:1202.6251}}\relax
\mciteBstWouldAddEndPuncttrue
\mciteSetBstMidEndSepPunct{\mcitedefaultmidpunct}
{\mcitedefaultendpunct}{\mcitedefaultseppunct}\relax
\EndOfBibitem
\bibitem{LHCb-PAPER-2013-018}
LHCb collaboration, R.~Aaij {\em et~al.},
  \ifthenelse{\boolean{articletitles}}{{\it {First observation of \CP violation
  in the decays of bottom strange mesons}},
  }{}\href{http://dx.doi.org/10.1103/PhysRevLett.110.221601}{Phys.\ Rev.\
  Lett.\  {\bf 110} (2013) 221601}, \href{http://arxiv.org/abs/1304.6173}{{\tt
  arXiv:1304.6173}}\relax
\mciteBstWouldAddEndPuncttrue
\mciteSetBstMidEndSepPunct{\mcitedefaultmidpunct}
{\mcitedefaultendpunct}{\mcitedefaultseppunct}\relax
\EndOfBibitem
\bibitem{LHCb-PAPER-2012-002}
LHCb collaboration, R.~Aaij {\em et~al.},
  \ifthenelse{\boolean{articletitles}}{{\it {Measurement of $b$-hadron
  branching fractions for two-body decays into charmless charged hadrons}},
  }{}\href{http://dx.doi.org/10.1007/JHEP10(2012)037}{JHEP {\bf 10} (2012) 37},
  \href{http://arxiv.org/abs/1206.2794}{{\tt arXiv:1206.2794}}\relax
\mciteBstWouldAddEndPuncttrue
\mciteSetBstMidEndSepPunct{\mcitedefaultmidpunct}
{\mcitedefaultendpunct}{\mcitedefaultseppunct}\relax
\EndOfBibitem
\bibitem{Lees:2013bb}
BaBar collaboration, J.~P. Lees {\em et~al.},
  \ifthenelse{\boolean{articletitles}}{{\it {Measurement of CP asymmetries and
  branching fractions in charmless two-body B-meson decays to pions and
  kaons}}, }{}\href{http://dx.doi.org/10.1103/PhysRevD.87.052009}{Phys.\ Rev.\
  {\bf D87} (2013) 052009}, \href{http://arxiv.org/abs/1206.3525}{{\tt
  arXiv:1206.3525}}\relax
\mciteBstWouldAddEndPuncttrue
\mciteSetBstMidEndSepPunct{\mcitedefaultmidpunct}
{\mcitedefaultendpunct}{\mcitedefaultseppunct}\relax
\EndOfBibitem
\bibitem{Adachi:2013mae}
Belle collaboration, I.~Adachi {\em et~al.},
  \ifthenelse{\boolean{articletitles}}{{\it {Measurement of the CP violation
  parameters in $B^0 \to \pi^+ \pi^-$ decays}},
  }{}\href{http://arxiv.org/abs/1302.0551}{{\tt arXiv:1302.0551}}\relax
\mciteBstWouldAddEndPuncttrue
\mciteSetBstMidEndSepPunct{\mcitedefaultmidpunct}
{\mcitedefaultendpunct}{\mcitedefaultseppunct}\relax
\EndOfBibitem
\bibitem{bib:hfagbase}
Heavy Flavor Averaging Group, Y.~Amhis {\em et~al.},
  \ifthenelse{\boolean{articletitles}}{{\it {Averages of $b$-hadron,
  $c$-hadron, and $\tau$-lepton properties as of early 2012}},
  }{}\href{http://arxiv.org/abs/1207.1158}{{\tt arXiv:1207.1158}}\relax
\mciteBstWouldAddEndPuncttrue
\mciteSetBstMidEndSepPunct{\mcitedefaultmidpunct}
{\mcitedefaultendpunct}{\mcitedefaultseppunct}\relax
\EndOfBibitem
\bibitem{Aaij:2013gta}
LHCb collaboration, R.~Aaij {\em et~al.},
  \ifthenelse{\boolean{articletitles}}{{\it {Measurement of the
  flavour-specific CP-violating asymmetry $a_{\rm sl}^s$ in $B_s^0$ decays}},
  }{}\href{http://arxiv.org/abs/1308.1048}{{\tt arXiv:1308.1048}}, {submitted
  to Phys. Lett. B}\relax
\mciteBstWouldAddEndPuncttrue
\mciteSetBstMidEndSepPunct{\mcitedefaultmidpunct}
{\mcitedefaultendpunct}{\mcitedefaultseppunct}\relax
\EndOfBibitem
\bibitem{LHCb-PAPER-2011-014}
LHCb collaboration, R.~Aaij {\em et~al.},
  \ifthenelse{\boolean{articletitles}}{{\it {Measurement of the effective
  $B^0_s\rightarrow K^+K^-$ lifetime}},
  }{}\href{http://dx.doi.org/10.1016/j.physletb.2011.12.058}{Phys.\ Lett.\
  {\bf B707} (2012) 349}, \href{http://arxiv.org/abs/1111.0521}{{\tt
  arXiv:1111.0521}}\relax
\mciteBstWouldAddEndPuncttrue
\mciteSetBstMidEndSepPunct{\mcitedefaultmidpunct}
{\mcitedefaultendpunct}{\mcitedefaultseppunct}\relax
\EndOfBibitem
\bibitem{LHCb-PAPER-2012-013}
LHCb collaboration, R.~Aaij {\em et~al.},
  \ifthenelse{\boolean{articletitles}}{{\it {Measurement of the effective
  $B_s^0 \to K^+ K^-$ lifetime}},
  }{}\href{http://dx.doi.org/10.1016/j.physletb.2012.08.033}{Phys.\ Lett.\
  {\bf B716} (2012) 393}, \href{http://arxiv.org/abs/1207.5993}{{\tt
  arXiv:1207.5993}}\relax
\mciteBstWouldAddEndPuncttrue
\mciteSetBstMidEndSepPunct{\mcitedefaultmidpunct}
{\mcitedefaultendpunct}{\mcitedefaultseppunct}\relax
\EndOfBibitem
\bibitem{Alves:2008zz}
LHCb collaboration, A.~A. Alves~Jr {\em et~al.},
  \ifthenelse{\boolean{articletitles}}{{\it {The \lhcb detector at the LHC}},
  }{}\href{http://dx.doi.org/10.1088/1748-0221/3/08/S08005}{JINST {\bf 3}
  (2008) S08005}\relax
\mciteBstWouldAddEndPuncttrue
\mciteSetBstMidEndSepPunct{\mcitedefaultmidpunct}
{\mcitedefaultendpunct}{\mcitedefaultseppunct}\relax
\EndOfBibitem
\bibitem{LHCb-DP-2012-003}
M.~Adinolfi {\em et~al.}, \ifthenelse{\boolean{articletitles}}{{\it
  {Performance of the \lhcb RICH detector at the LHC}},
  }{}\href{http://dx.doi.org/10.1140/epjc/s10052-013-2431-9}{Eur.\ Phys.\ J.\
  {\bf C73} (2013) 2431}, \href{http://arxiv.org/abs/1211.6759}{{\tt
  arXiv:1211.6759}}\relax
\mciteBstWouldAddEndPuncttrue
\mciteSetBstMidEndSepPunct{\mcitedefaultmidpunct}
{\mcitedefaultendpunct}{\mcitedefaultseppunct}\relax
\EndOfBibitem
\bibitem{LHCb-DP-2012-002}
A.~A. Alves~Jr {\em et~al.}, \ifthenelse{\boolean{articletitles}}{{\it
  {Performance of the LHCb muon system}},
  }{}\href{http://dx.doi.org/10.1088/1748-0221/8/02/P02022}{JINST {\bf 8}
  (2013) P02022}, \href{http://arxiv.org/abs/1211.1346}{{\tt
  arXiv:1211.1346}}\relax
\mciteBstWouldAddEndPuncttrue
\mciteSetBstMidEndSepPunct{\mcitedefaultmidpunct}
{\mcitedefaultendpunct}{\mcitedefaultseppunct}\relax
\EndOfBibitem
\bibitem{LHCb-DP-2012-004}
R.~Aaij {\em et~al.}, \ifthenelse{\boolean{articletitles}}{{\it {The \lhcb
  trigger and its performance in 2011}},
  }{}\href{http://dx.doi.org/10.1088/1748-0221/8/04/P04022}{JINST {\bf 8}
  (2013) P04022}, \href{http://arxiv.org/abs/1211.3055}{{\tt
  arXiv:1211.3055}}\relax
\mciteBstWouldAddEndPuncttrue
\mciteSetBstMidEndSepPunct{\mcitedefaultmidpunct}
{\mcitedefaultendpunct}{\mcitedefaultseppunct}\relax
\EndOfBibitem
\bibitem{BBDT}
V.~V. Gligorov and M.~Williams, \ifthenelse{\boolean{articletitles}}{{\it
  {Efficient, reliable and fast high-level triggering using a bonsai boosted
  decision tree}},
  }{}\href{http://dx.doi.org/10.1088/1748-0221/8/02/P02013}{JINST {\bf 8}
  (2013) P02013}, \href{http://arxiv.org/abs/1210.6861}{{\tt
  arXiv:1210.6861}}\relax
\mciteBstWouldAddEndPuncttrue
\mciteSetBstMidEndSepPunct{\mcitedefaultmidpunct}
{\mcitedefaultendpunct}{\mcitedefaultseppunct}\relax
\EndOfBibitem
\bibitem{Sjostrand:2006za}
T.~Sj\"{o}strand, S.~Mrenna, and P.~Skands,
  \ifthenelse{\boolean{articletitles}}{{\it {PYTHIA 6.4 physics and manual}},
  }{}\href{http://dx.doi.org/10.1088/1126-6708/2006/05/026}{JHEP {\bf 05}
  (2006) 026}, \href{http://arxiv.org/abs/hep-ph/0603175}{{\tt
  arXiv:hep-ph/0603175}}\relax
\mciteBstWouldAddEndPuncttrue
\mciteSetBstMidEndSepPunct{\mcitedefaultmidpunct}
{\mcitedefaultendpunct}{\mcitedefaultseppunct}\relax
\EndOfBibitem
\bibitem{LHCb-PROC-2010-056}
I.~Belyaev {\em et~al.}, \ifthenelse{\boolean{articletitles}}{{\it {Handling of
  the generation of primary events in \gauss, the \lhcb simulation framework}},
  }{}\href{http://dx.doi.org/10.1109/NSSMIC.2010.5873949}{Nuclear Science
  Symposium Conference Record (NSS/MIC) {\bf IEEE} (2010) 1155}\relax
\mciteBstWouldAddEndPuncttrue
\mciteSetBstMidEndSepPunct{\mcitedefaultmidpunct}
{\mcitedefaultendpunct}{\mcitedefaultseppunct}\relax
\EndOfBibitem
\bibitem{Lange:2001uf}
D.~J. Lange, \ifthenelse{\boolean{articletitles}}{{\it {The EvtGen particle
  decay simulation package}},
  }{}\href{http://dx.doi.org/10.1016/S0168-9002(01)00089-4}{Nucl.\ Instrum.\
  Meth.\  {\bf A462} (2001) 152}\relax
\mciteBstWouldAddEndPuncttrue
\mciteSetBstMidEndSepPunct{\mcitedefaultmidpunct}
{\mcitedefaultendpunct}{\mcitedefaultseppunct}\relax
\EndOfBibitem
\bibitem{Golonka:2005pn}
P.~Golonka and Z.~Was, \ifthenelse{\boolean{articletitles}}{{\it {PHOTOS Monte
  Carlo: a precision tool for QED corrections in $Z$ and $W$ decays}},
  }{}\href{http://dx.doi.org/10.1140/epjc/s2005-02396-4}{Eur.\ Phys.\ J.\  {\bf
  C45} (2006) 97}, \href{http://arxiv.org/abs/hep-ph/0506026}{{\tt
  arXiv:hep-ph/0506026}}\relax
\mciteBstWouldAddEndPuncttrue
\mciteSetBstMidEndSepPunct{\mcitedefaultmidpunct}
{\mcitedefaultendpunct}{\mcitedefaultseppunct}\relax
\EndOfBibitem
\bibitem{Allison:2006ve}
Geant4 collaboration, J.~Allison {\em et~al.},
  \ifthenelse{\boolean{articletitles}}{{\it {Geant4 developments and
  applications}}, }{}\href{http://dx.doi.org/10.1109/TNS.2006.869826}{IEEE
  Trans.\ Nucl.\ Sci.\  {\bf 53} (2006) 270}\relax
\mciteBstWouldAddEndPuncttrue
\mciteSetBstMidEndSepPunct{\mcitedefaultmidpunct}
{\mcitedefaultendpunct}{\mcitedefaultseppunct}\relax
\EndOfBibitem
\bibitem{Agostinelli:2002hh}
Geant4 collaboration, S.~Agostinelli {\em et~al.},
  \ifthenelse{\boolean{articletitles}}{{\it {Geant4: a simulation toolkit}},
  }{}\href{http://dx.doi.org/10.1016/S0168-9002(03)01368-8}{Nucl.\ Instrum.\
  Meth.\  {\bf A506} (2003) 250}\relax
\mciteBstWouldAddEndPuncttrue
\mciteSetBstMidEndSepPunct{\mcitedefaultmidpunct}
{\mcitedefaultendpunct}{\mcitedefaultseppunct}\relax
\EndOfBibitem
\bibitem{LHCb-PROC-2011-006}
M.~Clemencic {\em et~al.}, \ifthenelse{\boolean{articletitles}}{{\it {The \lhcb
  simulation application, \gauss: design, evolution and experience}},
  }{}\href{http://dx.doi.org/10.1088/1742-6596/331/3/032023}{{J.\ Phys.\ Conf.\
  Ser.\ } {\bf 331} (2011) 032023}\relax
\mciteBstWouldAddEndPuncttrue
\mciteSetBstMidEndSepPunct{\mcitedefaultmidpunct}
{\mcitedefaultendpunct}{\mcitedefaultseppunct}\relax
\EndOfBibitem
\bibitem{Breiman}
L.~Breiman, J.~H. Friedman, R.~A. Olshen, and C.~J. Stone, {\em Classification
  and regression trees}, Wadsworth international group, Belmont, California,
  USA, 1984\relax
\mciteBstWouldAddEndPuncttrue
\mciteSetBstMidEndSepPunct{\mcitedefaultmidpunct}
{\mcitedefaultendpunct}{\mcitedefaultseppunct}\relax
\EndOfBibitem
\bibitem{AdaBoost}
R.~E. Schapire and Y.~Freund, \ifthenelse{\boolean{articletitles}}{{\it A
  decision-theoretic generalization of on-line learning and an application to
  boosting}, }{}\href{http://dx.doi.org/10.1006/jcss.1997.1504}{Jour.\ Comp.\
  and Syst.\ Sc.\  {\bf 55} (1997) 119}\relax
\mciteBstWouldAddEndPuncttrue
\mciteSetBstMidEndSepPunct{\mcitedefaultmidpunct}
{\mcitedefaultendpunct}{\mcitedefaultseppunct}\relax
\EndOfBibitem
\bibitem{Aaij:2012mu}
LHCb collaboration, R.~Aaij {\em et~al.},
  \ifthenelse{\boolean{articletitles}}{{\it {Opposite-side flavour tagging of B
  mesons at the LHCb experiment}},
  }{}\href{http://dx.doi.org/10.1140/epjc/s10052-012-2022-1}{Eur.\ Phys.\ J.\
  {\bf C72} (2012) 2022}, \href{http://arxiv.org/abs/1202.4979}{{\tt
  arXiv:1202.4979}}\relax
\mciteBstWouldAddEndPuncttrue
\mciteSetBstMidEndSepPunct{\mcitedefaultmidpunct}
{\mcitedefaultendpunct}{\mcitedefaultseppunct}\relax
\EndOfBibitem
\bibitem{Baracchini:2005wp}
E.~Baracchini and G.~Isidori, \ifthenelse{\boolean{articletitles}}{{\it
  {Electromagnetic corrections to non-leptonic two-body B and D decays}},
  }{}\href{http://dx.doi.org/10.1016/j.physletb.2005.11.072}{Phys.\ Lett.\
  {\bf B633} (2006) 309}, \href{http://arxiv.org/abs/hep-ph/0508071}{{\tt
  arXiv:hep-ph/0508071}}\relax
\mciteBstWouldAddEndPuncttrue
\mciteSetBstMidEndSepPunct{\mcitedefaultmidpunct}
{\mcitedefaultendpunct}{\mcitedefaultseppunct}\relax
\EndOfBibitem
\bibitem{Albrecht:1989ga}
ARGUS collaboration, H.~Albrecht {\em et~al.},
  \ifthenelse{\boolean{articletitles}}{{\it {Search for $b \to s \gamma$ in
  exclusive decays of B mesons}},
  }{}\href{http://dx.doi.org/10.1016/0370-2693(89)91177-5}{Phys.\ Lett.\  {\bf
  B229} (1989) 304}\relax
\mciteBstWouldAddEndPuncttrue
\mciteSetBstMidEndSepPunct{\mcitedefaultmidpunct}
{\mcitedefaultendpunct}{\mcitedefaultseppunct}\relax
\EndOfBibitem
\bibitem{Cranmer:2000du}
K.~S. Cranmer, \ifthenelse{\boolean{articletitles}}{{\it {Kernel estimation in
  high-energy physics}},
  }{}\href{http://dx.doi.org/10.1016/S0010-4655(00)00243-5}{Comput.\ Phys.\
  Commun.\  {\bf 136} (2001) 198},
  \href{http://arxiv.org/abs/hep-ex/0011057}{{\tt arXiv:hep-ex/0011057}}\relax
\mciteBstWouldAddEndPuncttrue
\mciteSetBstMidEndSepPunct{\mcitedefaultmidpunct}
{\mcitedefaultendpunct}{\mcitedefaultseppunct}\relax
\EndOfBibitem
\bibitem{LHCb-PAPER-2013-002}
LHCb collaboration, R.~Aaij {\em et~al.},
  \ifthenelse{\boolean{articletitles}}{{\it {Measurement of \CP-violation and
  the $B^0_s$-meson decay width difference with $B_s^0\to J/\psi K^+K^-$ and
  $B_s^0 \to J/\psi\pi^+\pi^-$ decays}},
  }{}\href{http://dx.doi.org/10.1103/PhysRevD.87.112010}{Phys.\ Rev.\  {\bf
  D87} (2013) 112010}, \href{http://arxiv.org/abs/1304.2600}{{\tt
  arXiv:1304.2600}}\relax
\mciteBstWouldAddEndPuncttrue
\mciteSetBstMidEndSepPunct{\mcitedefaultmidpunct}
{\mcitedefaultendpunct}{\mcitedefaultseppunct}\relax
\EndOfBibitem
\bibitem{Aaij:2013mpa}
LHCb collaboration, R.~Aaij {\em et~al.},
  \ifthenelse{\boolean{articletitles}}{{\it {Precision measurement of the
  $B^0_s - \bar{B}^0_s$ oscillation frequency with the decay $B^0_s \to
  D^-_s\pi^+$}}, }{}\href{http://dx.doi.org/10.1088/1367-2630/15/5/053021}{New
  J.\ Phys.\  {\bf 15} (2013) 053021},
  \href{http://arxiv.org/abs/1304.4741}{{\tt arXiv:1304.4741}}\relax
\mciteBstWouldAddEndPuncttrue
\mciteSetBstMidEndSepPunct{\mcitedefaultmidpunct}
{\mcitedefaultendpunct}{\mcitedefaultseppunct}\relax
\EndOfBibitem
\bibitem{XIE}
LHCb collaboration, R.~Aaij {\em et~al.},
  \ifthenelse{\boolean{articletitles}}{{\it {Determination of the sign of the
  decay width difference in the $B^0_s$ system}},
  }{}\href{http://dx.doi.org/10.1103/PhysRevLett.108.241801}{Phys.\ Rev.\
  Lett.\  {\bf 108} (2012) 241801}, \href{http://arxiv.org/abs/1202.4717}{{\tt
  arXiv:1202.4717}}\relax
\mciteBstWouldAddEndPuncttrue
\mciteSetBstMidEndSepPunct{\mcitedefaultmidpunct}
{\mcitedefaultendpunct}{\mcitedefaultseppunct}\relax
\EndOfBibitem
\bibitem{Aaij:2012nt}
LHCb collaboration, R.~Aaij {\em et~al.},
  \ifthenelse{\boolean{articletitles}}{{\it {Measurement of the $B^0$--$\bar
  B^0$ oscillation frequency $\Delta m_d$ with the decays $B^0 \to D^- \pi^+$
  and $B^0 \to J/\psi K^{*0}$}},
  }{}\href{http://dx.doi.org/10.1016/j.physletb.2013.01.019}{Phys.\ Lett.\
  {\bf B719} (2013) 318}, \href{http://arxiv.org/abs/1210.6750}{{\tt
  arXiv:1210.6750}}\relax
\mciteBstWouldAddEndPuncttrue
\mciteSetBstMidEndSepPunct{\mcitedefaultmidpunct}
{\mcitedefaultendpunct}{\mcitedefaultseppunct}\relax
\EndOfBibitem
\bibitem{Beneke:2003zv}
M.~Beneke and M.~Neubert, \ifthenelse{\boolean{articletitles}}{{\it {QCD
  factorization for $B \to PP$ and $B \to PV$ decays}},
  }{}\href{http://dx.doi.org/10.1016/j.nuclphysb.2003.09.026}{Nucl.\ Phys.\
  {\bf B675} (2003) 333}, \href{http://arxiv.org/abs/hep-ph/0308039}{{\tt
  arXiv:hep-ph/0308039}}\relax
\mciteBstWouldAddEndPuncttrue
\mciteSetBstMidEndSepPunct{\mcitedefaultmidpunct}
{\mcitedefaultendpunct}{\mcitedefaultseppunct}\relax
\EndOfBibitem
\bibitem{DescotesGenon:2006wc}
S.~Descotes-Genon, J.~Matias, and J.~Virto,
  \ifthenelse{\boolean{articletitles}}{{\it {Exploring $B_{d,s} \to KK$ decays
  through flavor symmetries and QCD factorization}},
  }{}\href{http://dx.doi.org/10.1103/PhysRevLett.97.061801}{Phys.\ Rev.\ Lett.\
   {\bf 97} (2006) 061801}, \href{http://arxiv.org/abs/hep-ph/0603239}{{\tt
  arXiv:hep-ph/0603239}}\relax
\mciteBstWouldAddEndPuncttrue
\mciteSetBstMidEndSepPunct{\mcitedefaultmidpunct}
{\mcitedefaultendpunct}{\mcitedefaultseppunct}\relax
\EndOfBibitem
\bibitem{Chiang:2006ih}
C.-W. Chiang and Y.-F. Zhou, \ifthenelse{\boolean{articletitles}}{{\it {Flavor
  SU(3) analysis of charmless $B$ meson decays to two pseudoscalar mesons}},
  }{}\href{http://dx.doi.org/10.1088/1126-6708/2006/12/027}{JHEP {\bf 0612}
  (2006) 027}, \href{http://arxiv.org/abs/hep-ph/0609128}{{\tt
  arXiv:hep-ph/0609128}}\relax
\mciteBstWouldAddEndPuncttrue
\mciteSetBstMidEndSepPunct{\mcitedefaultmidpunct}
{\mcitedefaultendpunct}{\mcitedefaultseppunct}\relax
\EndOfBibitem
\end{mcitethebibliography}
